\documentclass[twocolumn]{aastex7}

\usepackage{placeins}
\usepackage{hyperref}
\usepackage{upgreek}
\usepackage{color}
\usepackage{amsmath}
\usepackage{graphicx}

\newcommand\note[1]{\textbf{\color{red}#1}}   
\newcommand\todo[1]{\textbf{\noindent \color{red} $\square$ #1}}
\renewcommand\note[1]{\hspace{-1sp}}            
\renewcommand\todo[1]{\hspace{-1sp}}

\shorttitle{KOINTREAU I: Two New Faint Companions To Taurus Stars}
\shortauthors{Walker et al.}

\begin{document}

\title{Keck Observations in the INfrared of Taurus and $\rho$ Oph Exoplanets And Ultracool dwarfs (KOINTREAU) I: A Planetary-Mass Companion and a Disk-Obscured Stellar Companion Discovered in Taurus}

\author[0000-0001-7062-815X]{Samuel A. U. Walker}
\affiliation{Institute for Astronomy at the University of Hawai`i at M\={a}noa, 2680 Woodlawn Drive, Hawai`i, HI 96822, USA}
\email{swalk@hawaii.edu}

\author[0000-0003-2232-7664]{Michael C. Liu}
\affiliation{Institute for Astronomy at the University of Hawai`i at M\={a}noa, 2680 Woodlawn Drive, Hawai`i, HI 96822, USA}
\email{mliu@ifa.hawaii.edu}

\author[0000-0002-8895-4735]{Dimitri Mawet}
\affiliation{Department of Astronomy, California Institute of Technology, Pasadena, CA 91125, USA}
\affiliation{Jet Propulsion Laboratory, California Institute of Technology, 4800 Oak Grove Dr., Pasadena, CA 91109, USA}
\email{dmawet@astro.caltech.edu}

\author{Charlotte Bond}
\affiliation{UK Astronomy Technology Centre, Blackford Hill, Edinburgh, Scotland, EH9 3HJ}
\email{charlotte.bond@stfc.ac.uk}

\author[0000-0002-8462-0703]{Mark Chun}
\affiliation{Institute for Astronomy at the University of Hawai`i at M\={a}noa, 2680 Woodlawn Drive, Hawai`i, HI 96822, USA}
\email{markchun@hawaii.edu}

\author[0000-0001-6301-896X]{Raquel A. Martinez}
\affiliation{Department of Physics \& Biophysics, University of San Diego, 5998 Alcal\'a Park, San Diego, CA 92110, USA}
\email{raquelmartinez@sandiego.edu}

\author[0000-0001-6041-7092]{Mark W. Phillips}
\affiliation{Institute for Astronomy at the University of Hawai`i at M\={a}noa, 2680 Woodlawn Drive, Hawai`i, HI 96822, USA}
\affiliation{Institute for Astronomy, University of Edinburgh, Blackford Hill, Edinburgh, Scotland, EH9 3HJ}
\email{mark.phillips@roe.ac.uk}

\author[0000-0001-5058-695X]{Jonathan P. Williams}
\affiliation{Institute for Astronomy at the University of Hawai`i at M\={a}noa, 2680 Woodlawn Drive, Hawai`i, HI 96822, USA}
\email{jw@hawaii.edu}

\author[0000-0002-3726-4881]{Zhoujian Zhang} 
\affiliation{Department of Physics \& Astronomy, University of Rochester, Rochester, NY 14627, USA}
\affiliation{Department of Astronomy \& Astrophysics, University of California, Santa Cruz, CA 95064, USA}
\email{zhangdirac@gmail.com}

\author[0000-0003-1698-9696]{Bin B. Ren}
\affiliation{P.E.S., Observatoire de la Côte d'Azur, Nice 06304, France}
\email{bin.ren@oca.eu}

\author[0000-0002-2805-7338]{Karl Stapelfeldt}
\affiliation{Jet Propulsion Laboratory, 4800 Oak Grove Drive, M/S 321-161, Pasadena, CA 91109}
\email{karl.r.stapelfeldt@jpl.nasa.gov}

\author[0000-0002-6879-3030]{Taichi Uyama}
\affiliation{California State University Northridge, 8111 Nordhoff Street, Northridge, CA 91330-8268}
\email{taichi.uyama.astro@gmail.com}

\author[0000-0003-0354-0187]{Nicole Wallack}
\affiliation{Carnegie Earth \& Planets Laboratory, 5241 Broad Branch Road, Washington, DC 20015}
\email{nwallack@carnegiescience.edu}


\begin{abstract}
\noindent
We present the first discoveries from Keck Observations in the INfrared of Taurus and $\rho$ Oph Exoplanets And Ultracool dwarfs (KOINTREAU), an adaptive optics imaging survey of young stars in the Taurus and $\rho$ Oph star-forming regions using the Keck infrared pyramid wavefront sensor (PyWFS).
We have found two faint ($\Delta K$$\sim$7~mag), wide-separation companions to two $\approx$3-Myr-old Taurus members. Relative astrometry for these systems show that both companions are bound to their host stars. We obtained near-infrared spectra of these companions using IRTF/SpeX ($R$$\sim$100) and Gemini/GNIRS ($R$$\sim$1000\nobreakdash-\hspace{0pt}2000), and combine these with photometry from our NIRC2 imaging, the Pan-STARRS survey, and \textit{Spitzer}/IRAC archival imaging to constrain their properties. One companion, KOINTREAU\nobreakdash-\hspace{0pt}1b (at a projected separation of 690 au), has an average near-IR spectral type of M9 $\pm$ 2, a gravity classification of \textsc{vl-g}, and a changing spectral type between the SpeX (M7) and GNIRS (L1) observations. We estimate this object's mass to be $10.6^{+2.5}_{-2.3}$ M$_{\rm Jup}$, making KOINTREAU\nobreakdash-\hspace{0pt}1b the fifth planetary-mass companion found in Taurus. The other companion, KOINTREAU\nobreakdash-\hspace{0pt}2b (projected separation 560 au), has a spectral type of M4.5 $\pm$ 1 but is $\approx4$ magnitudes underluminous relative to other Taurus stars of the same spectral type. We detect exceptionally strong He~I 1.083 $\upmu$m emission from this object, indicative of outflows driven by ongoing accretion, but with a conspicuous lack of accompanying H emission. We conclude that KOINTREAU\nobreakdash-\hspace{0pt}2b is a young star obscured by an edge-on disk and observed in scattered light. 
Finally, we derive a distortion solution for NIRC2 imaging which shows a 0.118$^\circ$ difference in position angle from the previous distortion solution.
\end{abstract}



\section{Introduction} \label{sec:intro}

By virtue of its sensitivity to massive ($\gtrsim$1 M$_{\rm Jup}$), wide-orbit ($\gtrsim$10 au) companions, direct imaging is able to probe a population of exoplanets and brown dwarfs that is infeasible to detect by other means \citep{bowler2016}. Additionally, the capture of photons directly emitted from such companions permits comparison with atmospheric and formation models independent of the light from the host star, unlike the study of transiting companions \citep[e.g.,][]{konopacky2013, sappey2025}.

Surveys of nearby young moving groups (YMGs; e.g., $\beta$ Pictoris and TW Hydrae) have produced a number of exoplanet discoveries \citep[e.g.,][]{marois2008, lagrange2009, carson2013, bowler2015, macintosh2015, derosa2023, franson2023, mesa2023}. YMG stars make attractive targets due to a combination of their proximity and relative youth, which conspire to ease the challenge of detecting exoplanets both by increasing the sky-projected distance between host star and companion for a given physical separation and by enhancing the contrast of the companion relative to its host star. As a result, the nearest YMGs have been extensively surveyed with adaptive optics to image brown dwarfs and exoplanets \citep[e.g.,][]{biller2013, galicher2016, nielsen2019, vigan2021}. Star-forming regions such as Taurus and $\rho$ Ophiuchus, while further away ($\approx$140 pc), are even younger ($\approx$1-5 Myr) than these YMGs and present a less thoroughly characterized frontier for discovery. Furthermore, the extreme youth of these regions means they provide a window into the earliest stages of wide-orbit planet and brown dwarf formation.

There have been several previous surveys of the Taurus star-forming region on a range of angular scales. On the largest scales, there have been efforts to construct a complete inventory of Taurus sources, primarily using surveys such as \textit{Gaia} \citep{gaiadr3} and the 2 Micron All Sky Survey \citep[2MASS;][]{cutri2003, skrutskie2006} and complemented by spectroscopic follow-up \citep[e.g.,][]{kraus2017, zhang2018, esplin2019, krolikowski2021}. These studies have provided key insights on the bulk properties of Taurus, including (but not limited to) that Taurus consists of a number of subgroups with ages ranging from $\approx$1 to 15-20 Myr \citep{kraus2017, zhang2018, kerr2021, krolikowski2021}; that the mass function of the younger and older Taurus subpopulations differ from one another, suggesting different formation histories \citep{zhang2018}; and that $\approx$70\% of Taurus stars with spectral types later than M3 host circumstellar disks \citep{esplin2019}.

On the smallest angular scales, there have been several imaging surveys using large-aperture telescopes with adaptive optics (AO) to inspect systems one by one, either with \citep[e.g.,][]{kraus2011, wallace2020} or without \citep[e.g.,][]{daemgen2015} the use of advanced techniques (coronagraphy or aperture masking) to aid in the detection of faint companions. These searches have constrained the overall occurrence rate of massive planets on wide orbits at young ages, e.g., \citet{wallace2020} finds that $<$30\% of Taurus stars host 2-13 M$_{\rm J}$ companions on orbits with semi-major axes from 10--500 au. In turn, these constraints inform the development of planet formation models.

To date, there are only four planetary-mass ($\lesssim$15 M$_{\rm Jup}$) companions to Taurus stars that have been directly imaged: DH Tau B \citep[spectral type M9;][]{itoh2005}, FU Tau B \citep[M9;][]{luhman2009futau}, 2M0441+2301 Bb \citep[L1;][]{bowler2015aabbab}, and 2MASS J0437 b \citep[mid-to-late L;][]{gaidos2022}. 
New discoveries would thus significantly increase the sample size of these objects.
There are also a number of isolated planetary-mass objects that have been found by mining wide-field imaging surveys \citep[e.g.,][]{zhang2018, esplin2019}, but the distances and ages of these objects are less well-constrained compared to bound companions of brighter, well-characterized stars.

In this paper we present the discovery of two new faint companions to the Taurus stars XEST 17-036 and XEST 13-010, the first published discoveries from Keck Observations in the INfrared of Taurus and $\rho$ Oph Exoplanets And Ultracool dwarfs (KOINTREAU), an AO imaging survey of young stars in Taurus and $\rho$ Oph using the Keck infrared pyramid wavefront sensor. We confirm these companions to be bound to their host stars by combining our relative astrometry of the companions from our imaging with \textit{Gaia} astrometry of their hosts. We characterize these companions by obtaining photometry (from NIRC2, Pan-STARRS and \textit{Spitzer}/IRAC) and spectroscopy (from IRTF/SpeX and Gemini/GNIRS). We show that the companion to XEST 17-036, hereinafter KOINTREAU\nobreakdash-\hspace{0pt}1b, has a late-M spectral type and is the fifth planetary-mass companion found in Taurus. We also show that KOINTREAU\nobreakdash-\hspace{0pt}2b, the companion to XEST 13-010, is a young early-to-mid-M dwarf occulted by an edge-on circumstellar disk.

\section{Target Stars}\label{sec:targs}

XEST 17-036, an M4 star \citep{jliu2021} in a $3.3\pm0.9$ Myr Taurus subcluster \citep{kerr2021}, was first found to be a Taurus member by \citet{scelsi2007} by matching X-ray data from the XMM-Newton Extended Survey of Taurus \citep{xest} with 2MASS data \citep{cutri2003, skrutskie2006} and verifying that the source agreed with the locus of known Taurus objects on $JHK_S$ color-magnitude diagrams. 
\textit{Gaia} parallax and proper motion data has subsequently validated this membership \citep{luhman2017, esplin2019}.
Additionally, \citet{luhman2009} found that this object exhibits an H$\alpha$ equivalent width of $6\pm1$\AA, which corroborates its youth and thus its Taurus membership \citep[see][Figure 3]{kraus2017}.
\citet{cody2022} detected a multi-periodic signal in the K2 lightcurve of this star, with variability on 9.62 and 4.55 day timescales, which they attributed to starspot modulation. The 9-day period is also found by \citet{chen2020} in their analysis of Zwicky Transient Facility (ZTF) photometry. \citet{bulger2014} analyzed Herschel/PACS images of XEST 17-036 and detected spatially extended emission out to $\approx$7$^{\prime\prime}$ in the 70 $\upmu$m band and $\approx$10$^{\prime\prime}$ in the 160 $\upmu$m band at the 3$\sigma$ level. They coupled this with \textit{Spitzer}/IRAC photometry to construct an SED of the star which identifies it as a Class III source. 
We estimated the mass of XEST 17-036 by converting its spectral type into an effective temperature using \citet{herczeg2014}, then linearly interpolating the BHAC15 evolutionary model grid \citep{bhac2015} in log(age), log(effective temperature), absolute 2MASS $K_S$ magnitude, and log(mass) to estimate a stellar mass given XEST 17-036's spectral type, age, and brightness \citep[corrected for extinction using $A_V$ from][]{kraus2017}. This gave a value of $0.26\pm0.03$ M$_{\odot}$, which is presented along with other properties of XEST 17-036 in Table \ref{tab:xest17}.

XEST 13-010 is a young \citep[$3.4\pm0.5$ Myr;][]{kerr2021} M3 star \citep[][]{luhman2006} observed as part of the XMM-Newton Extended Survey of Taurus \citep[XEST,][]{xest}. It was first identified as a Taurus member by \citet{luhman2006} on the basis of both its infrared excess emission and extinction ($A_V>$1 mag).
This Taurus membership was independently confirmed by \citet{scelsi2007} using 2MASS color-magnitude diagrams in conjunction with XEST survey data, and has been reaffirmed using \textit{Gaia} data \citep{luhman2017, esplin2019}.
\citet{rebull2020}, \citet{roggero2021} and \citet{cody2022} all used Kepler/K2 data to find that XEST 13-010 exhibits periodic 3.9-day variability, which they attributed to rotational modulation from starspots. This is confirmed by \citet{chen2020} using ZTF data. \citet{roggero2021} and \citet{cody2022} also detected aperiodic dipping, which implies that the star is accreting material from a circumstellar disk. \citet{rebull2020} classified XEST 13-010 as a Class II source based on the shape of its SED, which provides additional evidence as to the presence of a circumstellar disk. \citet{akeson2019} used SMA observations of XEST 13-010 made by \citet{andrews2013} to estimate that this disk has a total mass of $1.7^{+0.3}_{-0.5}$~M$_{\rm Jup}$; they also calculate the host star to be $0.29^{+0.09}_{-0.05}$~M$_{\odot}$. Key properties of this star are reported in Table \ref{tab:xest13}.
\begin{deluxetable}{lcc}
\centering
\tabletypesize{\small}
\tablecaption{XEST 17-036 system properties\label{tab:xest17}}
\tablehead{\colhead{Property} & \colhead{Value}& \colhead{Ref.}}
\startdata
\multicolumn{3}{c}{XEST 17-036}\\
\hline
RA (J2000)&04 33 26.22& (2)\\
Dec. (J2000)&+22 45 29.3& (2)\\
$\mu_{\rm RA}$ (mas yr$^{-1}$) & $8.9 \pm 0.1$ & (2)\\
$\mu_{\rm Dec.}$ (mas yr$^{-1}$) & $-16.5 \pm 0.1$ & (2)\\
$\pi$ (mas) & $	6.3 \pm 0.1$ & (2)\\
Distance (pc) & $159 \pm 3$ & (2)\\
Spectral type & M4 & (3)\\
Mass (M$_{\odot}$) & $0.26\pm0.03$ & (1)\\
Age (Myr) & $3.3\pm0.9$ & (7) \\
$T_{\rm eff}$ (K) & $3551\pm59$ & (4)\\
log g (cm s$^{-2}$) & $3.44\pm0.05$ & (4)\\
{[Fe/H]} (dex) & $-0.15\pm0.01$ & (4)\\
$v\sin{i}$ (km s$^{-1}$) & $7.7\pm0.5$ & (5)\\
$A_V$ (mag) & $5.3\pm0.3$ & (6)\\
\hline
\multicolumn{3}{c}{KOINTREAU-1b}\\
\hline
Projected separation (au) & $690\pm10$ & (1)\\
Mass (M$_{\rm Jup}$) & $10.6^{+2.5}_{-2.3}$ & (1)\\
Spectral type & M9$\pm2$ \textsc{vl-g} & (1)\\
$A_V$ from photometry (mag) & $10\pm2$ & (1)\\
$A_V$ from spectra (mag) & 6.8 & (1)\\
\enddata
\tablerefs{
(1) This work;
(2) \citet{gaiadr3};
(3) \citet{jliu2021};
(4) \citet{jonsson2020};
(5) \citet{kounkel2019};
(6) \citet{kraus2017};
(7) \citet{kerr2021}
}
\end{deluxetable}
\begin{deluxetable}{lcc}
\centering
\tabletypesize{\small}
\tablecaption{XEST 13-010 system properties\label{tab:xest13}}
\tablehead{\colhead{Property} & \colhead{Value}& \colhead{Ref.}}
\startdata
\multicolumn{3}{c}{XEST 13-010}\\
\hline
RA (J2000)&04 29 36.06& (2)\\
Dec. (J2000)&+24 35 55.6& (2)\\
$\mu_{\rm RA}$ (mas yr$^{-1}$) & $8.64 \pm 0.05$ & (2)\\
$\mu_{\rm Dec.}$ (mas yr$^{-1}$) & $-20.64 \pm 0.04$ & (2)\\
$\pi$ (mas) & $7.63 \pm 0.04$ & (2)\\
Distance (pc) & $131.1 \pm 0.6$ & (2)\\
Spectral type & M3 & (3)\\
Mass (M$_{\odot}$) & $0.29^{+0.09}_{-0.05}$ & (7)\\
Age (Myr) & $3.4\pm0.5$ & (8) \\
$T_{\rm eff}$ (K) & $3622\pm60$ & (4)\\
log g (cm s$^{-2}$) & $3.38\pm0.05$ & (4)\\
{[Fe/H]} (dex) & $-0.05\pm0.01$ & (4)\\
$v\sin{i}$ (km s$^{-1}$) & $24.6\pm2.7$ & (5)\\
$A_V$ (mag) & $5.21\pm0.35$ & (6)\tablenotemark{a}\\
\hline
\multicolumn{3}{c}{KOINTREAU-2b}\\
\hline
Projected separation (au) & $561\pm3$ & (1)\\
Spectral type & M4.5$\pm1$ & (1)\\
$A_V$ from photometry (mag) & $6.8\pm0.7$ & (1)\\
$A_V$ from spectrum (mag) & 4.9 & (1)\\
\enddata
\tablenotetext{a}{Calculated from $A_J$ assuming the \citet{cardelli1989} extinction law.}
\tablerefs{
(1) This work;
(2) \citet{gaiadr3};
(3) \citet{luhman2006};
(4) \citet{jonsson2020};
(5) \citet{lopez2021};
(6) \citet{luhman2006};
(7) \citet{akeson2019};
(8) \citet{kerr2021};
}
\end{deluxetable}
\section{Observations} \label{sec:obs}
\subsection{Keck/NIRC2 Imaging} \label{sec:imaging}

We obtained adaptive optics (AO) imaging of XEST 17-036 and XEST 13-010 with the Keck Near-InfraRed Camera 2 (Keck/NIRC2), positioned behind the adaptive optics bench on the left Nasmyth platform of the 10m-aperture Keck II telescope. We used both the Shack-Hartmann wavefront sensor \citep[SHWFS;][]{wizinowich2000} with a laser guide star \citep[LGS;][]{wizinowich2006, vandam2006} and the infrared pyramid wavefront sensor \citep[PyWFS;][]{pywfsjatis} in natural guide star (NGS) mode. These observations are summarized in Table \ref{tab:nirc2}. Our observing procedure was to obtain a set of short exposures, used to register the unsaturated position of the host star, then a set of longer exposures (typically $3$-$5$ images of $60$s each) to detect any faint companions. We did not dither as part of our observations. We took data in the Mauna Kea Observatories \citep[MKO;][]{simons2002,tokunaga2002} $K$-band at each epoch, and obtained MKO $YJH$ data during our December 2023 observing run. In addition to our science data, we also took dome images in each filter with the flat lamp on and off, and dark frames with the shutter closed and the same exposure times as our science frames. These were median combined to produce master flat and dark frames. We then subtracted the corresponding master dark frame from our data, divided our dark-subtracted data by the relevant master flat frame, removed cosmic rays using \texttt{LACosmic} \citep{lacosmic}, and summed individual exposures together to form image stacks for each exposure time. We found one faint candidate companion around each of XEST 17-036 and XEST 13-010 by inspection of the stacked long-exposure images. These companions are hereinafter referred to as KOINTREAU\nobreakdash-\hspace{0pt}1b and KOINTREAU\nobreakdash-\hspace{0pt}2b respectively.

For each epoch, we centroided the positions of the central star and candidate companions using the individual short- and long-exposure images, respectively. We adopted the weighted mean and associated error on each set of measurements as the measured position of each star and companion. We then converted these positions into astrometry of the companion relative to the host star, correcting for NIRC2 distortion using the \citet{service2016} distortion correction and the 0.118$^\circ$ shift in position angle relative to the \citeauthor{service2016} correction that we find from our own analysis (Appendix \ref{ap:distortion}).

To determine the flux ratios of the companions, we performed aperture photometry on the target stars in the short-exposure images and their companions in the long-exposure images. 
The sky background was estimated using the median of a 100-pixel-wide border around the edge of each image and subtracted before measuring any photometry. Additionally, the deep-exposure images were median-subtracted in successive annuli around the host star to remove the light from the host star when measuring the brightness of the companion.
We chose our photometry aperture size independently for each set of observations by finding the aperture size that maximized the signal-to-noise ratio of the resulting contrast measurement (the radii of these apertures ranged from 11-66 pixels, reflecting the varying image quality of the observations).
The summed aperture counts for the host star were calculated from each short-exposure image that comprised the stacked image, and the same was done for the companion from each long-exposure image. We took the mean of the host star and companion counts from the set of images at each epoch, adopting the standard error on this set of measurements as our errors.
The ratio of the counts per second of the companion to that of the host star was then used to calculate the contrast of each companion for each set of observations. These values are presented in Table \ref{tab:nirc2}. 

There is statistically significant inter-epoch variability in the measured $K$-band contrasts for both systems. Since we only measure the brightness of the companion relative to its host, we are unable to determine whether this variability is due to changes in the host star, the companion, or both.  We adopt the weighted mean of our $K$-band contrasts going forward, with an error given by the weighted root mean square -- these values are $\Delta K=6.9\pm0.1$ mag for KOINTREAU\nobreakdash-\hspace{0pt}1b and $\Delta K=6.7\pm0.1$ mag for KOINTREAU\nobreakdash-\hspace{0pt}2b.

\begin{figure*}
     \centering
     \includegraphics[width=0.49\linewidth]{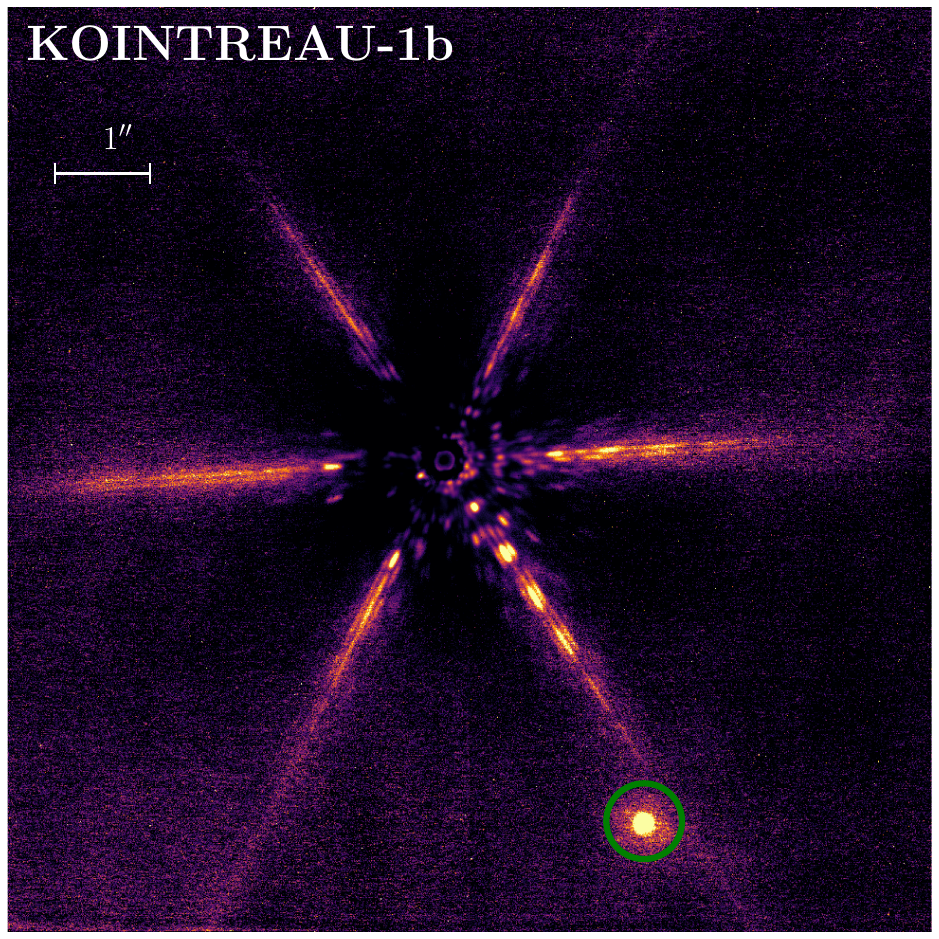}
    \centering
     \includegraphics[width=0.49\linewidth]{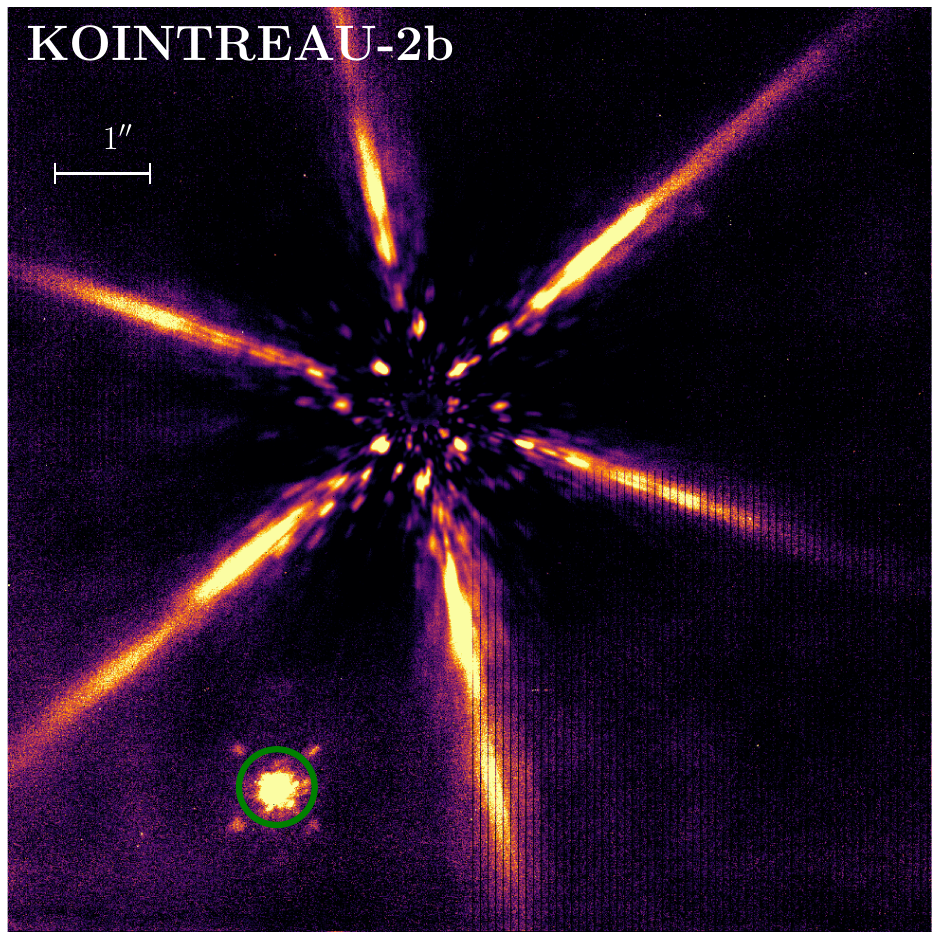}
     \caption{Example Keck/NIRC2 signal-to-noise maps from our deep-exposure 10$^{\prime\prime}$ $K$-band image stacks of each companion. These maps are computed by subtracting the mean and normalizing by the standard deviation in successive annuli around the host star. The companions are circled in green in each case. We note some electronic noise in the lower right quadrant of the shown epoch (December 2023) of KOINTREAU\nobreakdash-\hspace{0pt}2b data, but our measurements are unaffected by this.}
    \label{fig:nirc2}
\end{figure*}

We took the 2MASS magnitudes of the host star, transformed these to the MKO system \citep{skrutskie2006} and added these to our contrasts to determine the magnitudes of the candidate companions in each filter. These values are given in Table \ref{tab:xest17phot} and Table \ref{tab:xest13phot} for KOINTREAU\nobreakdash-\hspace{0pt}1b and KOINTREAU\nobreakdash-\hspace{0pt}2b, respectively.

\subsection{IRTF/SpeX Prism Spectra}

We obtained near-infrared (0.7--2.5~$\upmu$m) spectra of KOINTREAU\nobreakdash-\hspace{0pt}1b using SpeX, the medium-resolution 0.7-5.3 $\upmu$m spectrograph mounted at the Cassegrain focus of the 3.2m-aperture NASA InfraRed Telescope Facility (IRTF) \citep{2003PASP..115..362R}. We operated SpeX in its Prism mode ($R\sim100$) on the nights of UT 2020 November 27 and UT 2020 December 30 in $\approx0.9\arcsec$ seeing conditions. Ninety 120-second frames were taken in the prism mode with the $0.8\arcsec \times 15\arcsec$ slit and the standard ABBA nodding pattern. The airmass of our target was less than 1.25 during our observations. To minimize the contaminating light from its primary star, we fixed the slit position angle (PA) at $300^{\circ}$, perpendicular to the position angle of KOINTREAU\nobreakdash-\hspace{0pt}1b relative to its host. We note that the blurring caused by differential chromatic refraction (DCR) between 0.9--2.45 $\upmu$m is $\lesssim0.17\arcsec$ at airmass 1.25, which is less than our chosen slit size and the typical seeing atop Maunakea, thus DCR should have a negligible impact on our observations, despite not observing at the parallactic angle.
An A0V telluric standard star, HD~35036, was observed contemporaneously with KOINTREAU\nobreakdash-\hspace{0pt}1b with $<$0.12 airmass difference. Data were reduced using Spextool \citep{2004PASP..116..362C}. Our final science spectrum is the weighted mean and its formal uncertainty of the spectra from each night. This spectrum has a signal-to-noise ratio (SNR) of 20 per pixel near the peak of the $J$ band (1.28--1.32~$\upmu$m).

Additionally, we obtained spectra of KOINTREAU\nobreakdash-\hspace{0pt}2b with SpeX/Prism over UT 2024 March 20 and 21 with $\approx0.8\arcsec$ seeing on each night. 85 180-second frames were taken in the prism mode with the $0.5\arcsec \times 15\arcsec$ slit and the standard ABBA nodding pattern. The airmass of our target varied from 1.2 to 2 during our observations. We fixed the slit position angle perpendicular to the PA of the companion, as with KOINTREAU\nobreakdash-\hspace{0pt}1b. At airmass 2, the DCR blur size is $\lesssim0.38^{\prime\prime}$, so our observations will not be significantly affected by this. Our standard star for these observations was HD 1718. Our final reduced, stacked science spectrum has a signal-to-noise ratio of 50 per pixel near the peak of the $J$ band.

\subsection{Gemini/GNIRS Spectra}

We observed KOINTREAU\nobreakdash-\hspace{0pt}1b with the Gemini North InfraRed Spectrograph \citep[Gemini/GNIRS;][]{gnirs1, gnirs2} on the 8.1m-aperture Gemini North telescope using queue mode, collecting a total of 252 minutes of usable data on UT dates 2024 October 7, 2024 October 13, 2024 October 14, 2024 November 22 and 2024 December 6 (program IDs GN-2024B-Q-228 and GN-2024B-Q-317).
Data were collected using the cross-dispersed mode (which permits simultaneous 0.8--2.5 $\upmu$m spectral coverage) with the short blue camera, the $32~\mathrm{l}/\mathrm{mm}$ grating and $0.675\,\arcsec$ slit ($R$$\sim$800). The slit position angle was fixed perpendicular to the position angle of the candidate companion to minimize contamination from the host star. All science data were taken at airmasses $<$1.2. The blurring caused by DCR at airmass 1.2 between wavelengths of 0.9--2.45 $\upmu$m is $0.145^{\prime\prime}$, less than the size of our slit and the size of Maunakea seeing, so our data should not have been substantively affected by this. The exposures were dithered in an ABBA pattern along the slit. We observed the standard stars HIP 10559 ($V=5.25\,$mag), HIP 17704 ($V=6.8\,$mag) and HIP 29909 ($V=7.5\,$mag). 

KOINTREAU\nobreakdash-\hspace{0pt}2b was observed by Gemini/GNIRS in queue mode, collecting a total of 252 minutes of usable data (program IDs GN-2023B-Q-233 and GN-2024B-Q-317). Data were collected using the cross-dispersed mode of the short blue camera with the $32~\mathrm{l}/\mathrm{mm}$ grating and $0.3\,\arcsec$ slit ($R$$\sim$1800) on UT 2023 December 27 and UT 2024 January 23 and the $0.675\,\arcsec$ slit ($R$$\sim$800) on UT 2024 December 3 and UT 2024 December 19, using the same procedure as above. All science data were taken at airmasses $<$1.5, for which the relevant DCR blur size is $0.244^{\prime\prime}$, again less than the size of both our selected slits and MKO seeing. We observed the standard stars HIP 17588 ($V=6.4\,$mag) and HIP 29881 ($V=6.57\,$mag) for telluric correction, using the same instrument setup and dither pattern as for KOINTREAU\nobreakdash-\hspace{0pt}1b.

The Gemini/GNIRS spectra are affected by striping caused by the detector controller, noticeable in the raw frames as dark or bright stripes with an 8-pixel periodicity in one or more quadrants of the detector. To remove these artifacts we use the \texttt{CleanIR} python script provided by Gemini\footnote{\url{https://www.github.com/andrewwstephens/cleanir}}, which performs full frame pattern removal and quadrant leveling. The data are then reduced using the open-source Python package \texttt{PypeIt} \citep{bochanski2009, bernstein2015, prochaska2020a, prochaska2020b}, following a similar setup as described in \citet{phillips2024}. Briefly, we used ten night-time flats taken with a quartz-halogen lamp as the calibration frames required by \texttt{PypeIt} for order edge tracing and the pixel efficiency correction. The wavelength solution was determined from OH skylines in the science frames, which is preferable to using the Ar lamp frames due to flexure. \texttt{PypeIt} achieved Poisson-limited sky subtraction \citep{prochaska2020b} and then performed an optimal extraction to generate 1D science spectra. Flux calibration and telluric correction were performed using an infrared sensitivity function generated from the observations of the standard star. After extraction, \texttt{PypeIt} rebinned the spectra onto a common logarithmic wavelength grid, facilitating the co-addition of multiple exposures and the combination of overlapping echelle orders. For more information on the algorithms used in the \texttt{PypeIt} data reduction pipeline, we refer the reader to \citet{bochanski2009}, \citet{bernstein2015}, and the \texttt{PypeIt} online documentation. 

We formed our final Gemini/GNIRS science spectra of both targets by taking the weighted mean and formal uncertainty of each set of spectra.

\subsection{Pan-STARRS photometry}

Pan-STARRS 1 is a 1.8m-aperture wide-field optical survey telescope used to image the entire night sky visible from Hawai`i in five bandpasses $g, r, i, z,$ and $y$, which collectively cover a wavelength range of 430-990 nm. The primary data products are stacked images made from the multiple exposures taken of each point on the sky, a catalog of photometry constructed from these image stacks, and the individual exposures (``warps'') that constitute the stacks \citep{ps1, magnier2020}.

\begin{figure*}
    \centering
         \includegraphics[width=\linewidth]{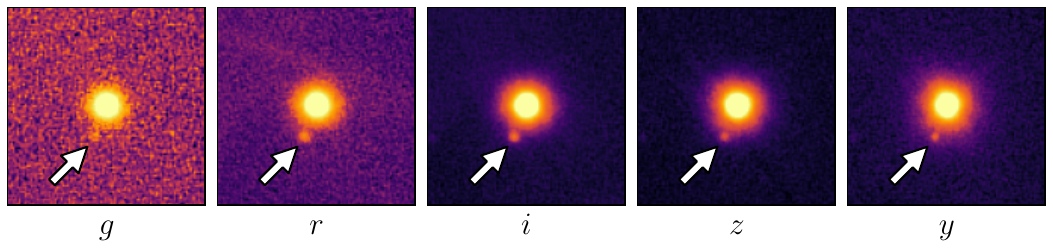}
         \caption{Pan-STARRS $grizy$ 13$^{\prime\prime}$ images of the XEST 13-010 system. The arrow in each image indicates KOINTREAU\nobreakdash-\hspace{0pt}2b.}
         \label{fig:xest13_ps1}
\end{figure*}

KOINTREAU\nobreakdash-\hspace{0pt}2b is visible by eye in all of the $grizy$ stacks (Figure~\ref{fig:xest13_ps1}), although KOINTREAU\nobreakdash-\hspace{0pt}1b is not. However, the Pan-STARRS DR1 catalog only reports $ri$ magnitudes for KOINTREAU\nobreakdash-\hspace{0pt}2b, and these values may be inaccurate given KOINTREAU\nobreakdash-\hspace{0pt}2b's proximity to its significantly brighter host star. Thus, we extracted our own photometry of KOINTREAU\nobreakdash-\hspace{0pt}2b from the Pan-STARRS image data. We downloaded 2$^\prime$ cutouts centered on XEST 13-010 from each warp in each filter and then subtracted the median radial profile of the host star XEST 13-010. We then measured the flux of KOINTREAU\nobreakdash-\hspace{0pt}2b in a 5-pixel aperture, as well as the flux of 7 well-detected unsaturated reference stars that have photometry in the Pan-STARRS catalog. We repeated this for each warp (with the exception of two $i$-band warps that were taken in $>$2$^{\prime\prime}$ seeing), and computed the flux ratio of KOINTREAU\nobreakdash-\hspace{0pt}2b relative to each of the reference stars in each warp. We took the mean and standard error of the flux ratios of KOINTREAU\nobreakdash-\hspace{0pt}2b relative to each reference star for each filter's set of warps. We then computed $grizy$ magnitudes for KOINTREAU\nobreakdash-\hspace{0pt}2b by combining our computed contrasts with the catalog magnitudes for each comparison star. Finally, we took the weighted average (and uncertainty) of the resulting set of magnitudes as the $grizy$ magnitudes for KOINTREAU\nobreakdash-\hspace{0pt}2b (Table \ref{tab:xest13phot}).

\begin{rotatetable*}
\begin{deluxetable*}{lcccccccccc}

\centering

\tabletypesize{\footnotesize}
\tablecaption{NIRC2 AO Imaging Summary\label{tab:nirc2}}
\tablehead{\colhead{Target} & \colhead{Date} & \colhead{Filter} & Camera & \colhead{Shallow} & \colhead{Deep} & \colhead{AO mode/} & \colhead{\textbf{Separation}} & \colhead{\textbf{PA}} & \colhead{Contrast} & \colhead{FWHM}\\
\colhead{} & \colhead{(UT)} & \colhead{(MKO)} & & \colhead{exp. time\tablenotemark{a}} & \colhead{ exp. time\tablenotemark{a}} & \colhead{WFS\tablenotemark{b}} & \colhead{(mas)} & \colhead{(deg.)} & \colhead{($\Delta$mag.)} & \colhead{(mas)\tablenotemark{c}}}
\startdata
KOINTREAU-1b& 2019-12-16& $K$ & Narrow & $3\times20\times1$ & $4\times1\times60$ & NGS/Py & $4342\pm4$ & $208.44\pm0.03$ & $6.4\pm0.1$&$36\pm4$\\
& 2020-10-09 & $K$ & Narrow & $4\times10\times1$ & $4\times1\times60$    & NGS/Py & $4333\pm3$ & $208.50\pm0.06$ &$6.88\pm0.02$&$28\pm1$\\
& 2021-10-27 & $K$ & Narrow & $5\times10\times0.36$ & $5\times1\times60$ & NGS/Py & $4335\pm3$ & $208.38\pm0.05$ & $6.82\pm0.06$&$29\pm1$\\
& 2022-12-12 & $K$ & Wide   & $6\times5\times1$ & $11\times1\times60$    & NGS/Py & $4347\pm17$ & $208.5\pm0.3$ &$6.7\pm0.1$&$156\pm20$\\
& 2023-12-03 & $K$ & Narrow & $5\times2\times5$ & $5\times1\times60$     & NGS/Py & $4333\pm12$ & $208.3\pm0.2$ & $7.3\pm0.1$&$78\pm9$\\
& 2023-12-16 & $Y$ & Narrow & $6\times2\times5$ & $3\times1\times120$    & LGS/SH &$-$ &$-$ &$>6.0$\tablenotemark{d}&$69\pm1$\\
& 2023-12-16 & $J$ & Narrow & $6\times12\times1.5$ & $5\times1\times60$  & LGS/SH & $4330\pm15$ & $208.4\pm0.2$ &$7.90\pm0.03$&$69\pm8$\\
& 2023-12-16 & $H$ & Narrow & $6\times25\times0.4$ & $5\times1\times60$  & LGS/SH & $4336\pm12$ & $208.4\pm0.2$ &$7.40\pm0.01$&$47\pm10$\\
\hline
KOINTREAU-2b& 2019-12-16 & $K$ & Narrow & $3\times100\times0.18$ & $3\times1\times60$ & NGS/Py & $4283\pm2$ & $158.80\pm0.02$ &$6.55\pm0.04$&$23\pm0.3$\\
& 2020-11-26 & $K$ & Wide & $9\times200\times0.05$ & $5\times1\times20$   & NGS/Py &$4282\pm15$ &$158.8\pm0.1$ &$5.9\pm0.2$&$133\pm32$\\
& 2023-12-07 & $K$ & Narrow & $5\times50\times0.187$ & $5\times1\times60$ & LGS/SH &$4281\pm8$ &$158.9\pm0.1$ &$6.73\pm0.02$&$30\pm2$\\
& 2023-12-16 & $Y$ & Narrow & $6\times25\times0.4$ & $5\times1\times60$   & LGS/SH &$4291\pm15$ &$158.8\pm0.2$ &$6.5\pm0.1$&$45\pm1$\\
& 2023-12-16 & $J$ & Narrow & $6\times25\times0.4$ & $5\times1\times60$   & LGS/SH &$4291\pm15$ &$158.8\pm0.2$ &$6.74\pm0.02$&$53\pm9$\\
& 2023-12-16 & $H$ & Narrow & $6\times50\times0.17$ & $5\times1\times60$  & LGS/SH &$4285\pm11$ &$158.9\pm0.2$ &$6.75\pm0.01$&$40\pm9$\\
\enddata
\tablenotetext{a}{Number of exposures $\times$ Number of coadds $\times$ Integration time per coadd (s)}
\tablenotetext{b}{Natural guide star mode with the Pyramid wavefront sensor (NGS/Py) or laser guide star mode with the Shack-Hartmann wavefront sensor (LGS/SH)}
\tablenotetext{c}{Calculated from the radial profile of the host star in the shallow-exposure stack}
\tablenotetext{d}{Lower limit calculated from contrast curve of observation stack.}
\end{deluxetable*}
\end{rotatetable*}

\begin{deluxetable}{lcc}
\centering
\tabletypesize{\small}
\tablecaption{XEST 17-036 system photometry\label{tab:xest17phot}}
\tablehead{\colhead{Band} & \colhead{Value (mag.)}& \colhead{Ref.}}
\startdata
\multicolumn{3}{c}{XEST 17-036}\\
\hline
Pan-STARRS $g$ &$20.60\pm0.03$ & (5)\\
Pan-STARRS $r$ &$18.188\pm0.009$ & (5)\\
Pan-STARRS $i$ &$15.930\pm0.008$ & (5)\\
Pan-STARRS $z$ &$14.701\pm0.008$ & (5)\\
Pan-STARRS $y$ &$13.933\pm0.003$ & (5)\\
\textit{Gaia} $G$& $16.496\pm0.002$ & (2)\\
\textit{Gaia} $BP$ & $19.49\pm0.06$ & (2)\\
\textit{Gaia} $RP$ & $14.971\pm0.005$ & (2)\\
$J_{\rm MKO}$& $11.69\pm0.03$ & (3,4)\tablenotemark{a}\\
$H_{\rm MKO}$& $10.52\pm0.03$ & (3,4)\tablenotemark{a}\\
$K_{\rm MKO}$& $9.88\pm0.02$ & (3,4)\tablenotemark{a}\\
IRAC $[3.6]$ &$9.53\pm0.02$ & (1)\\
IRAC $[4.5]$ &$9.39\pm0.02$ & (1)\\
IRAC $[5.8]$ &$9.26\pm0.02$ & (1)\\
IRAC $[8.0]$ &$9.31\pm0.02$ & (1)\\
\hline
\multicolumn{3}{c}{KOINTREAU-1b}\\
\hline
$J_{\rm MKO}$ & $19.59\pm0.04$ & (1)\\
$H_{\rm MKO}$ & $17.92\pm0.04$ & (1)\\
$K_{\rm MKO}$ & $16.7\pm0.1$ & (1)\\
IRAC $[3.6]$ &$15.53\pm0.03$ & (1)\\
IRAC $[4.5]$ &$15.07\pm0.03$ & (1)\\
IRAC $[5.8]$ &$15.1\pm0.1$ & (1)\\
IRAC $[8.0]$ &$14.5\pm0.2$ & (1)\\
\enddata
\tablenotetext{a}{Transformed from 2MASS magnitudes using relations given in  \citet{skrutskie2006}.}
\tablerefs{
(1) This work;
(2) \citet{gaiadr3};
(3) \citet{cutri2003};
(4) \citet{skrutskie2006};
(5) \citet{ps1}
}
\end{deluxetable}

\begin{deluxetable}{lcc}
\centering
\tabletypesize{\small}
\tablecaption{XEST 13-010 system photometry\label{tab:xest13phot}}
\tablehead{\colhead{Band} & \colhead{Value (mag.)}& \colhead{Ref.}}
\startdata
\multicolumn{3}{c}{XEST 13-010}\\
\hline
Pan-STARRS $g$ &$17.61\pm0.05$ & (5)\\
Pan-STARRS $r$ &$15.691\pm0.005$ & (5)\\
Pan-STARRS $i$ &$14.17\pm0.02$ & (5)\\
Pan-STARRS $z$ &$13.22\pm0.01$ & (5)\\
Pan-STARRS $y$ &$12.81\pm0.03$ & (5)\\
\textit{Gaia} $G$& $14.78\pm0.01$ & (2)\\
\textit{Gaia} $BP$ & $16.78\pm0.03$ & (2)\\
\textit{Gaia} $RP$ & $13.45\pm0.04$ & (2)\\
$J_{\rm MKO}$ & $10.65\pm0.03$ & (3,4)\tablenotemark{a}\\
$H_{\rm MKO}$ & $9.43\pm0.03$ & (3,4)\tablenotemark{a}\\
$K_{\rm MKO}$ & $8.60\pm0.02$ & (3,4)\tablenotemark{a}\\
IRAC $[3.6]$ &$8.00\pm0.02$ & (1)\\
IRAC $[4.5]$ &$7.67\pm0.02$ & (1)\\
IRAC $[5.8]$ &$7.38\pm0.02$ & (1)\\
IRAC $[8.0]$ &$6.96\pm0.02$ & (1)\\
\hline
\multicolumn{3}{c}{KOINTREAU-2b}\\
\hline
Pan-STARRS $g$ &$23.2\pm0.6$ & (1)\\
Pan-STARRS $r$ &$21.0\pm0.2$ & (1)\\
Pan-STARRS $i$ &$20.00\pm0.09$ & (1)\\
Pan-STARRS $z$ &$19.4\pm0.2$ & (1)\\
Pan-STARRS $y$ &$18.8\pm0.1$ & (1)\\
\textit{Gaia} $G$& $20.91\pm0.03$ & (2)\\
\textit{Gaia} $BP$ & $22.02\pm0.06$ & (2)\\
\textit{Gaia} $RP$ & $19.5\pm0.2$ & (2)\\
$J_{\rm MKO}$ & $17.39\pm0.04$ & (1)\\
$H_{\rm MKO}$ & $16.18\pm0.04$ & (1)\\
$K_{\rm MKO}$ & $15.3\pm0.1$ & (1)\\
IRAC $[3.6]$ &$14.30\pm0.04$ & (1)\\
IRAC $[4.5]$ &$13.62\pm0.02$ & (1)\\
IRAC $[5.8]$ &$13.91\pm0.14$ & (1)\\
IRAC $[8.0]$ &$>13.87$ & (1)\\
\enddata
\tablenotetext{a}{Transformed from 2MASS magnitudes using relations given in  \citet{skrutskie2006}.}
\tablerefs{
(1) This work;
(2) \citet{gaiadr3};
(3) \citet{cutri2003};
(4) \citet{skrutskie2006};
(5) \citet{ps1}
}
\end{deluxetable}

\begin{figure*}
     \centering
     \includegraphics[width=\linewidth]{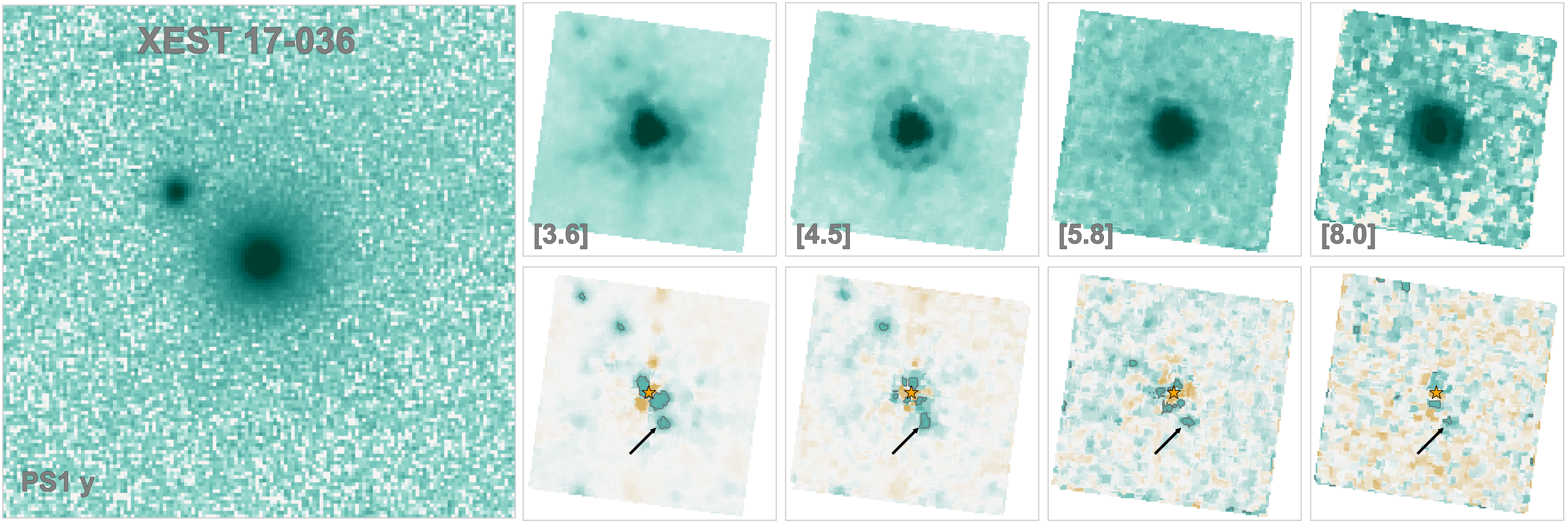}    

     \vfill

    \centering
     \includegraphics[width=\linewidth]{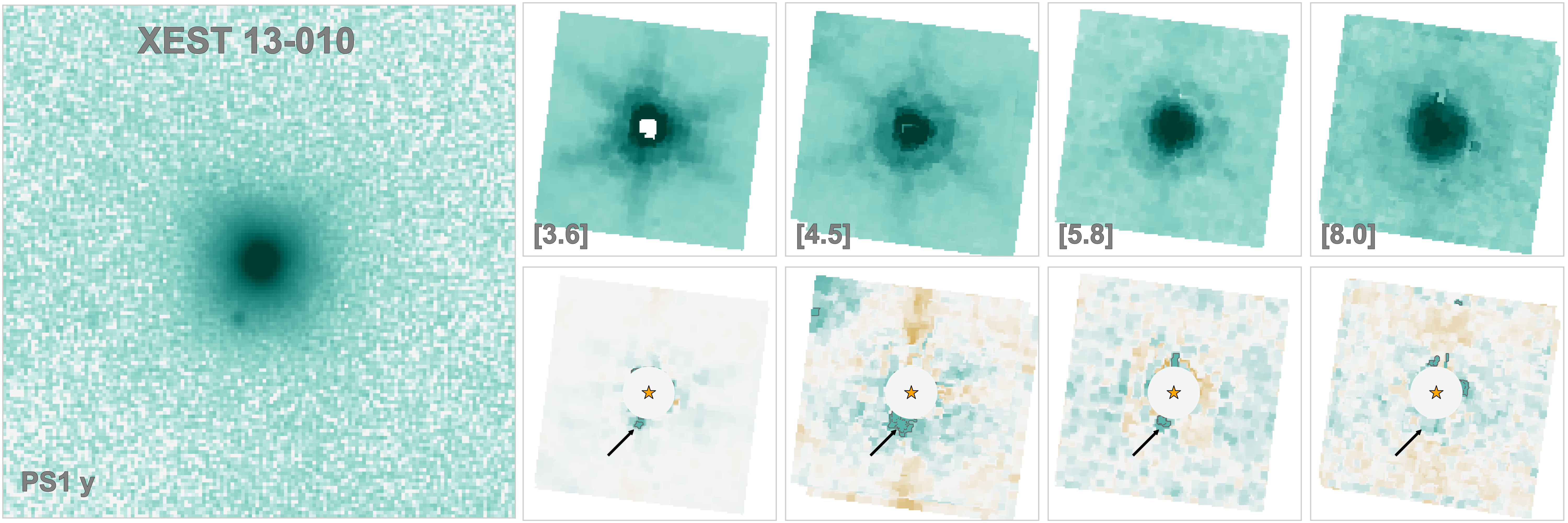}
     \caption{The results of our \textit{Spitzer}/IRAC PSF fitting procedure for XEST~17-036 (top) and XEST~13-010 (bottom), with the positions of each candidate companion indicated with a black arrow. Despite not being visible by eye in the catalog images, fitting and subtracting PSFs of the host star reveals the faint companions detected in our Keck/NIRC2 imaging.}
    \label{fig:spitzer}
\end{figure*}

\subsection{Spitzer/IRAC photometry}

The \textit{Spitzer Space Telescope} was an infrared space telescope that operated in its primary science mode from launch in 2003 until its liquid helium coolant ran out in 2009. The InfraRed Array Camera (IRAC) was one of the instruments on board \textit{Spitzer}, capable of simultaneously imaging a 5$^{\prime}$ patch of sky with a $1.2^{\prime\prime}$ pixel scale in four channels centered on 3.6 $\upmu$m, 4.5 $\upmu$m, 5.8 $\upmu$m and 8.0 $\upmu$m \citep{irac}.

Observations of XEST 17-036 and XEST 13-010 that were re-analyzed for this work appear in 4 IRAC programs, with exposure times of either 0.4 or 10.4 s. XEST 17-036 observations were taken as part of Program ID 37 (PI: G.~Fazio) on UT 2005 February 20, and as part of Program ID 30816 (PI: D.~Padgett) on UT 2007 April 3. XEST 13-010 IRAC observations were taken as part of Program ID 173 (PI: N.~Evans) on UT 2004 September 8, but only in [4.5] and [8.0]. Observations of XEST 13-010 in all four IRAC channels were executed as part of Program ID 3584 (PI: D.~Padgett) on UT 2005 February 24 and UT 2005 February 25.

We worked with IRAC's cryogenic-phase corrected basic calibrated data (CBCD) and uncertainty (CBUNC) files with exposure times of 10.4 s, avoiding mosaics due to the complicated nature of the IRAC PSF.
All data were reduced with the \textit{Spitzer} Science Center software pipeline version S18.25.0.

The angular resolution of IRAC makes it challenging to distinguish either KOINTREAU\nobreakdash-\hspace{0pt}1b or KOINTREAU\nobreakdash-\hspace{0pt}2b from their host stars by eye. Thus, we used the framework described in \cite{martinez2019} to model the PSFs of the system components in the IRAC images. To briefly summarize this procedure, we used the point-response function \citep[or effective PSF;][]{hoffman05} developed by the Spitzer science team to generate model PSFs at any position on the IRAC detector. We then fit a two-source PSF model in each image, performing an MCMC analysis using a Metropolis–Hastings algorithm with Gibbs sampling. The PSF model is described by seven parameters: the x-pixel coordinate of the primary centroid (x), y-pixel coordinate of the primary centroid (y), image background (b), peak pixel value of the primary (n), projected separation ($\rho$), position angle (PA), and contrast ($\Delta$m). In addition, image pixel values greater than $80\%$ of the saturation limit were masked. We used our high-precision NIRC2 astrometry measurements of the companions as priors in our Markov Chain Monte Carlo (MCMC) fits.

The MCMC analysis was conducted in two stages to determine image-specific parameters (x, y, b) separately from system-specific parameters (n, $\rho$, PA, $\Delta$m). We ran two MCMC chains with 140,000 total steps, discarding the first 10\% of each chain as “burn-in”. The weighted average median (x, y)-centroid, $\rho$, PA, and $\Delta$m generated by the MCMC fit were used to create individual PSF models of each system component from which aperture photometry using a 10$^{\prime\prime}$ radius was measured.
XEST 17-036 has a bright nearby neighbor with flux that could potentially influence the results of the pipeline fit. For these images, we generated a PSF model as described above and subtracted the nearby neighbor from each individual IRAC image before processing them through the pipeline.

Adopting this approach for both systems gave us the brightness of each host star and companion in each IRAC channel, although we were only able to constrain an upper limit to the brightness of KOINTREAU\nobreakdash-\hspace{0pt}2b in [8.0]. These data are presented in Table \ref{tab:xest17phot} and Table \ref{tab:xest13phot} for the XEST 17-036 and XEST 13-010 systems, respectively.

\section{Results}

\begin{figure*}
    \centering
    \includegraphics[width=.85\linewidth]{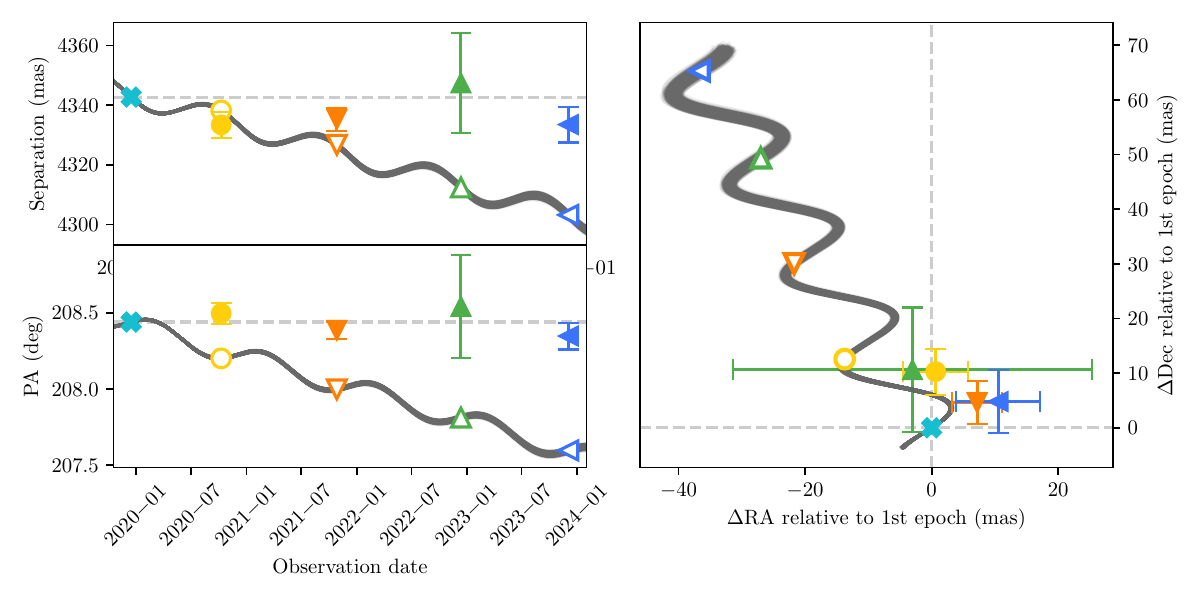}
    \caption{Relative astrometry of KOINTREAU\nobreakdash-\hspace{0pt}1b. The gray curve shows 1000 draws from a model of the predicted movement of a background object relative to the companion's first epoch detection (the blue `x' in each panel).
    The hollow markers indicate where the companion would appear if it were a background object at each epoch, and the filled markers indicate the measured position of the companion in the corresponding epoch. Note that measurements are presented relative to the first epoch, hence first-epoch measurement errors are incorporated into the displayed errorbars for subsequent epochs. Over a baseline of four years, the change in the companion's sky motion, separation and position angle relative to its host star are consistent with zero.}
    \label{fig:xest17astrometry}
\end{figure*}

\begin{figure*}
    \centering
    \includegraphics[width=.85\linewidth]{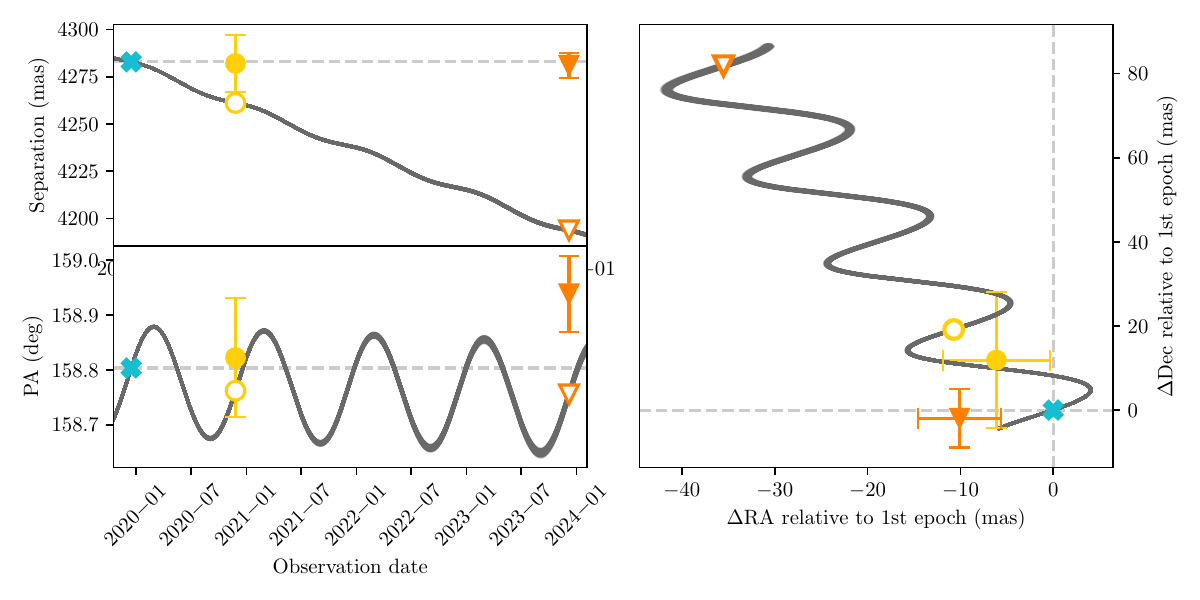}
    \caption{Relative astrometry of KOINTREAU\nobreakdash-\hspace{0pt}2b, the imaged companion to XEST 13-010, demonstrating that the two are physically associated. See Figure~\ref{fig:xest17astrometry} for full description.}
    \label{fig:xest13astrometry}
\end{figure*}

\subsection{Common Proper Motion}

We combined the relative astrometry data measured from our Keck/NIRC2 images with \textit{Gaia} DR3 proper motion and parallax data \citep{gaiadr3} to ascertain if the candidate companions exhibit proper motion common with their host stars. \textit{Gaia} data describes how the host star moves across the sky between epochs. The opposite of this motion thus describes the movement of a stationary background object relative to the star, which is the null hypothesis for our observed companions. If the data are inconsistent with this null hypothesis, we then conclude that the companions are physically associated with their respective host stars. The results of this analysis are plotted in Figure~\ref{fig:xest17astrometry} for KOINTREAU\nobreakdash-\hspace{0pt}1b, the candidate companion to XEST 17-036, and in Figure~\ref{fig:xest13astrometry} for KOINTREAU\nobreakdash-\hspace{0pt}2b, the candidate companion to XEST 13-010. The measured movements of both companions relative to their first-epoch positions are inconsistent in sky motion, separation, and position angle with the static background null hypothesis. We verify this by computing the reduced chi-squared statistic for both the background and comoving hypotheses for each object and comparing them using the Bayes factor, following \citet{bowler2013}. For KOINTREAU\nobreakdash-\hspace{0pt}1b, $\chi^2_{\nu, {\rm background}}=32.2$, compared to $\chi^2_{\nu, {\rm comoving}}=2.57$ (4 degrees of freedom), indicating log(Bayes factor) of 25.7; for KOINTREAU\nobreakdash-\hspace{0pt}2b $\chi^2_{\nu, {\rm background}}=58.5$ and $\chi^2_{\nu, {\rm comoving}}=0.83$ (2 degrees of freedom), giving log(Bayes factor) of 25.1. For both companions the comoving hypothesis is decisively preferred.

\subsection{Spectral Analysis}
\label{sec:spts}

\subsubsection{KOINTREAU\nobreakdash-\hspace{0pt}1b: SpeX/Prism}

We compared our IRTF SpeX/Prism spectrum of KOINTREAU\nobreakdash-\hspace{0pt}1b with published SpeX/Prism and SpeX/SXD spectra of the very-low-gravity (\textsc{vl-g}, i.e. young) M and L standards recommended by \citet{allersandliu2013} to estimate its spectral type and extinction. We convolved the SpeX/SXD data ($R$$\sim$750) with a Gaussian kernel and resampled it at the spectral resolution of our SpeX/Prism spectrum ($R$$\sim$100).
Using $\chi^2$ minimization we fit two parameters, $A_V$ and a normalization factor, with $A_{\lambda}$ calculated from $A_V$ using the \citet{cardelli1989} extinction law assuming $R=3.1$. We use data from 0.9--2.45~$\upmu$m for fitting, as data in this range consistently has an SNR $>5$. We find that TWA 8B, with a spectral type of M6 \textsc{vl-g} \citep{allers2009}, provides the best fit to our data with $A_V=7.1$ mag (Figure~\ref{fig:xest17spec}), a little larger than the $A_V$ of $5.3\pm0.3$ mag reported for the host star by \citet{kraus2017} based on its $r^\prime-K$ color. We note that fixing $A_V$ at the host star value produces a fit 1-2 subtypes later but of worse quality.
We dereddened the companion's spectrum using our best-fit extinction and then measured the spectral type using the \citet{allersandliu2013} spectral indices. These give an average spectral type of M6.5, with Monte Carlo methods providing an estimated error of $\pm0.9$ subtypes on this measurement. The \citet{allersandliu2013} indices also classify this object as having a gravity classification of \textsc{vl-g}, with gravity scores of 2, 2, and 2 from the FeH$_z$, KI$_J$ and H$_{\rm cont}$ indices respectively.

\subsubsection{KOINTREAU\nobreakdash-\hspace{0pt}1b: Gemini/GNIRS}

We compared our Gemini/GNIRS spectrum of KOINTREAU\nobreakdash-\hspace{0pt}1b with these same \citet{allersandliu2013} M and L standards, following the approach described above, except we convolve all standard spectra to SpeX/Prism resolution for consistency (if not observed at this resolution), then convolve our GNIRS spectrum to Prism resolution to match this. We mask out regions of low sky transmission in the GNIRS data from 1.35--1.49~$\upmu$m and 1.8--1.95~$\upmu$m after convolution.

\begin{figure*}[t]
    \centering
    \includegraphics[width=\linewidth]{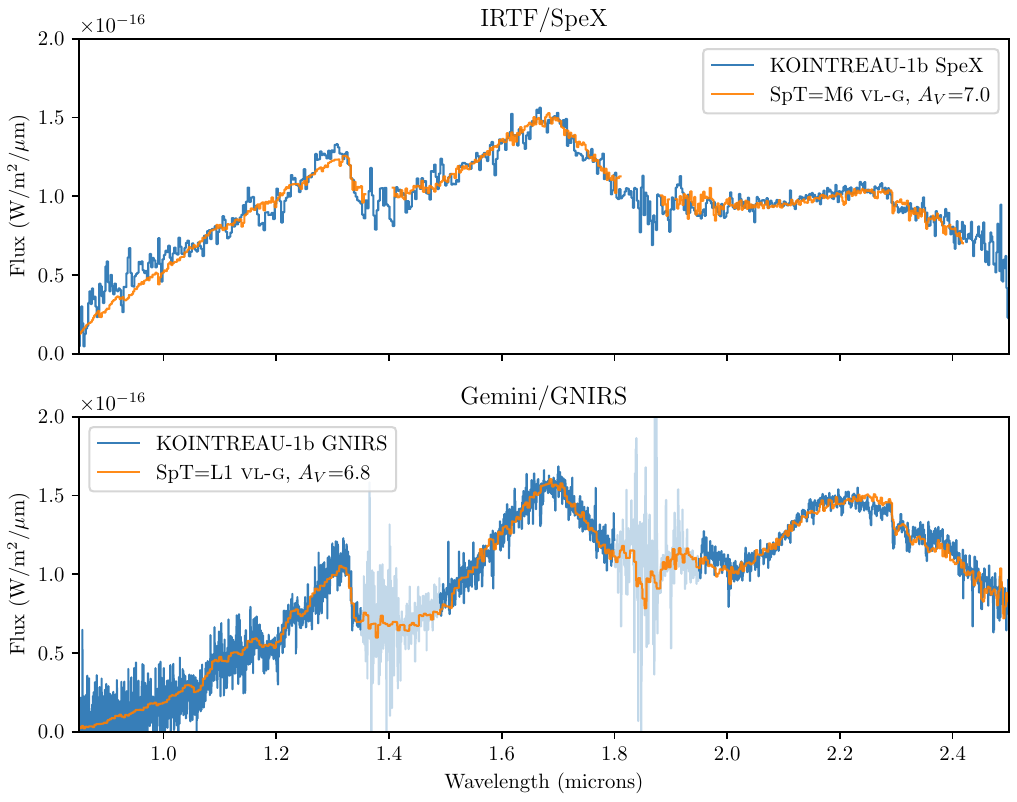}
    \caption{Spectra of KOINTREAU\nobreakdash-\hspace{0pt}1b. Overplotted are the two best-fitting spectral \citet{allersandliu2013} standards for each spectrum, which have been reddened and scaled to match our spectra, which themselves are flux-matched to our NIRC2 $H$-band photometry. The transparent regions of the data were unused in our analysis due to low sky transmission, and the faint blue shading around each spectrum indicates the measurement errors. For the IRTF/SpeX Prism spectrum, the best-fitting spectral type is M6, and the Gemini/GNIRS data has a best-fitting spectral type of L1.}
    \label{fig:xest17spec}
\end{figure*}

We find that 2MASS J05184616-2756457, with a spectral type of L1 \textsc{vl-g} \citep{allersandliu2013}, provides the best fit to our data with $A_V=6.8$ mag (Figure~\ref{fig:xest17spec}), which is greater than the $A_V$ of $5.3\pm0.3$ mag for the host star. Again, fixing $A_V$ at the host star value produces a poorer-quality fit that is 1-2 subtypes later than the best fit.
We dereddened our spectrum using the best-fit extinction, downsampled it to the resolution of SpeX/Prism data, 
and then measured its spectral type using the \citet{allersandliu2013} spectral indices, which gave an average spectral type of L0 $\pm$ 0.1. The \citet{allersandliu2013} indices also classify this object as \textsc{vl-g}, with gravity scores of 2, 2, 2, and 2 from the FeH, VO$_Z$, and H$_{\rm cont}$ indices and alkali line equivalent width measurements respectively. This agrees with the SpeX/Prism results and confirms XEST 17-036's youth.

\begin{figure*}[t]
    \centering
    \includegraphics[width=\linewidth]{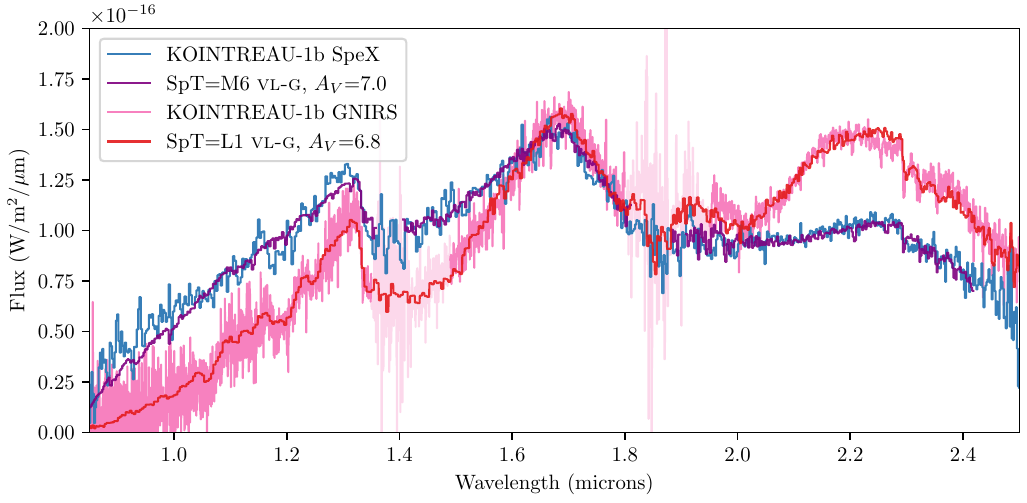}
    \caption{Direct comparison of IRTF/SpeX and Gemini/GNIRS spectra of KOINTREAU\nobreakdash-\hspace{0pt}1b, flux-matched to our NIRC2 $H$-band photometry. Each best-fit standard spectrum is also plotted, reddened and scaled to match the respective science spectrum. The spectral slope of the two datasets are clearly discrepant.}
    \label{fig:xest17speccomp}
\end{figure*}

KOINTREAU\nobreakdash-\hspace{0pt}1b's Gemini/GNIRS spectrum has a color ($J-K=2.8$ mag) which differs significantly from its 2020B IRTF/SpeX Prism spectrum ($J-K=2.1$ mag), as shown in Figure~\ref{fig:xest17speccomp}. We discuss the physical implications of this in Section~\ref{sec:xest17interpretation}.
Averaging our measurements from our SpeX/Prism and GNIRS spectra, we adopt a final spectral type of M9 $\pm$ 2 for this object.

\subsubsection{KOINTREAU\nobreakdash-\hspace{0pt}2b: SpeX/Prism}

We initially followed the same fitting approach with our IRTF SpeX/Prism spectrum of KOINTREAU\nobreakdash-\hspace{0pt}2b as we used with KOINTREAU\nobreakdash-\hspace{0pt}1b, additionally excluding from the fit the prominent He~I 1.08 $\upmu$m emission line. However, we found that none of the \citet{allersandliu2013} standard spectra provided a good match to our data -- the best fit had a spectral type of M5, but was a poor visual match to the data.

\begin{figure*}
    \centering
    \includegraphics[width=\linewidth]{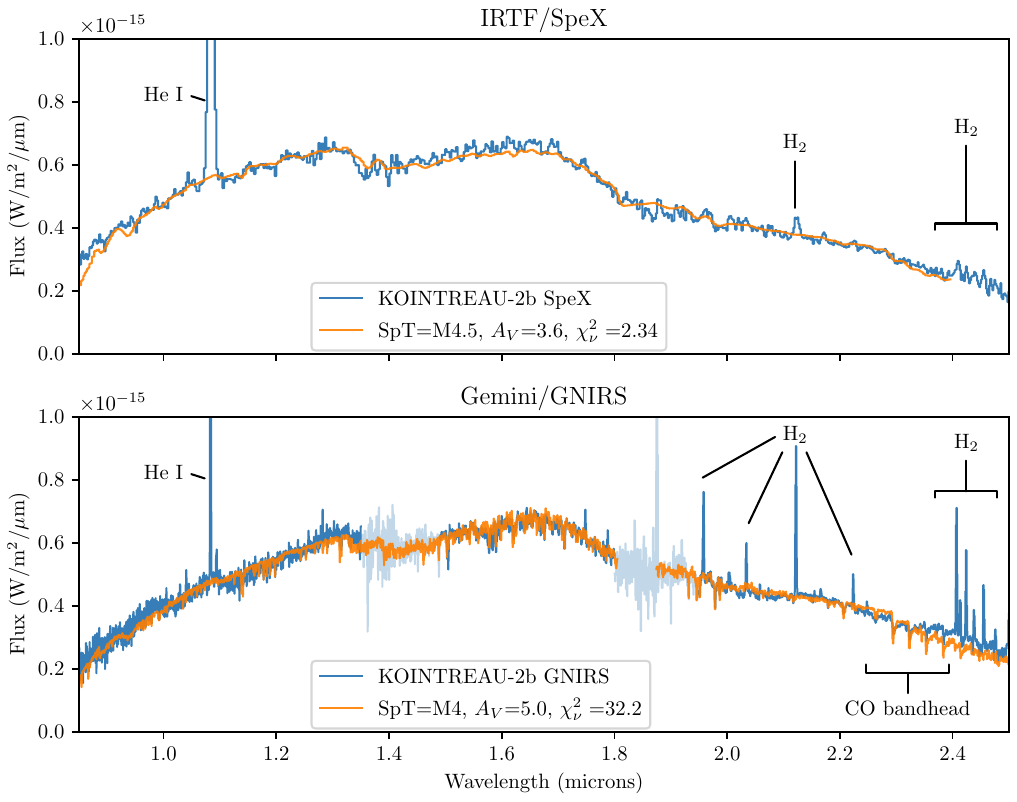}
    \caption{Spectra of KOINTREAU\nobreakdash-\hspace{0pt}2b. Overplotted are the two best-fitting giant$+$dwarf combination spectra from our analysis, which have been reddened and scaled to match our spectra, which are flux-matched to our NIRC2 $H$-band photometry. The transparent regions of the data were unused in our analysis due to low sky transmission, and the faint blue shading around each spectrum indicates the measurement errors. In both cases, the best-fitting spectral type is M4.5. In our Gemini/GNIRS spectrum the peak flux of the He~I line is 1.040$\pm$0.003$\times$10$^{-14}$ W/m$^2$/$\upmu$m, over an order of magnitude brighter than the continuum.}
    \label{fig:xest13spec}
\end{figure*}

As M5 is the earliest spectral type in the \citet{allersandliu2013} spectral standards, we expanded our spectral fitting approach to earlier spectral types using K and M stars from the IRTF SpeX Spectral Library \citep{rayner2009}.
To generate spectra for young pre-main-sequence stars, we follow the approach of \citet{luhman1999} in normalizing the IRTF SpeX Spectral Library spectra of giants and dwarfs of the same spectral type and averaging them. We degrade the spectral resolution of these resultant spectra to match the resolution of SpeX/Prism data. This set of standards produced better fits, as judged both by eye and by $\chi^2_\nu$.
Our best fit (Figure~\ref{fig:xest13spec}) is an M4.5 spectrum formed by combining HD 204585 (giant) and Gl 268AB (dwarf) with $A_V=3.6$ mag (Figure~\ref{fig:xest13spec}), which is lower than the $A_V=5.21\pm0.35$ mag that \citet{luhman2006} reports for the host star.

\subsubsection{KOINTREAU\nobreakdash-\hspace{0pt}2b: Gemini/GNIRS}

We also fit our hybrid giant-dwarf IRTF SpeX Spectral Library spectra described above to our Gemini/GNIRS spectrum of KOINTREAU\nobreakdash-\hspace{0pt}2b, excluding the H$_2$ 1.95 $\upmu$m, 2.03 $\upmu$m and 2.12 $\upmu$m emission lines from the fit.

Our best fit obtained in this manner (Figure~\ref{fig:xest13spec}) is the same M4.5 spectrum that best fit the SpeX/Prism data albeit with $A_V=4.9$ mag, which is consistent with the $A_V=5.21\pm0.35$ mag that \citet{luhman2006} reports for the host star.

\begin{figure}
    \centering
    \includegraphics[width=\linewidth]{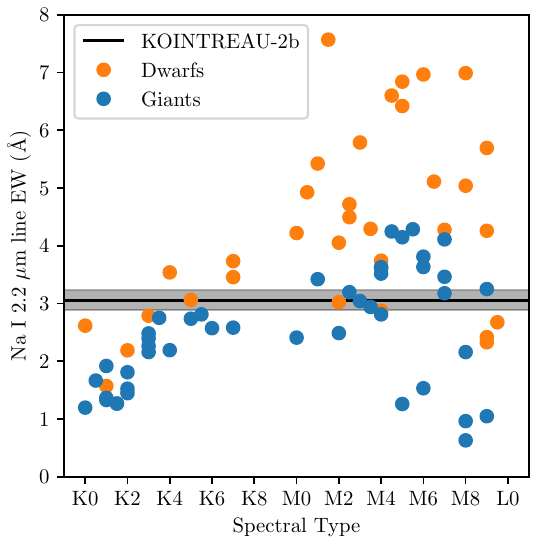}
    \caption{Na I 2.22 $\upmu$m equivalent width calculated from our Gemini/GNIRS spectrum of KOINTREAU\nobreakdash-\hspace{0pt}2b, plotted alongside equivalent widths calculated in the same manner for K and M stars from the IRTF SpeX Spectral Library. The shaded region indicates the error on KOINTREAU\nobreakdash-\hspace{0pt}2b's EW measurement, which is consistent with the spectral type we derive from our spectral fitting procedure. Errors on SpeX Spectral Library equivalent width measurements are equal to or smaller than the marker size.}
    \label{fig:naew}
\end{figure}

In addition, we deredden our science spectrum using the $A_V=4.9$ mag from our spectral fitting and then use the spectral type-equivalent width relations presented in \citet{rayner2009} to verify the spectral type for KOINTREAU\nobreakdash-\hspace{0pt}2b from our spectral fitting. By eye, only one of these lines has a sufficient SNR to reliably measure an equivalent width, the Na I 2.22 $\upmu$m doublet. Due to the lower SNR of our spectrum compared to those in \citet{rayner2009}, we compute equivalent widths over a smaller wavelength range than the one they use (2.204--$2.212$ $\upmu$m for us, compared to the \citeauthor{rayner2009} range of 2.185--$2.230$ $\upmu$m). We estimate the continuum flux across this region by fitting a straight line to the data from 2.192--2.198 $\upmu$m and 2.213--2.22 $\upmu$m. Using these ranges, we calculate equivalent widths for our science spectrum and the IRTF SpeX Spectral Library standard stars with spectral types from K0 to L0. We estimate errors on each equivalent width measurement by perturbing the observed fluxes using their errors in a Monte Carlo fashion 10,000 times and calculating the standard deviation of the resulting distribution of equivalent widths. The resulting measurements are presented in Figure~\ref{fig:naew}, indicating an early-to-mid M spectral type, which is in good agreement with our spectral fitting results.

The spectral type we find from our template fitting analysis (M4.5) is too early for the \citet{allersandliu2013} spectral indices to apply. However, we note that our best-fit spectra were found by mimicking the spectra of young low-gravity dwarfs, which indicates that KOINTREAU\nobreakdash-\hspace{0pt}2b also has low surface gravity, as expected given its Taurus membership.

\subsection{Extinction from photometry}
\label{subsub:photextinction}

We can see from our photometry and spectroscopy that both KOINTREAU\nobreakdash-\hspace{0pt}1b and KOINTREAU\nobreakdash-\hspace{0pt}2b vary in brightness over time. Given our measured spectral types of KOINTREAU\nobreakdash-\hspace{0pt}1b and KOINTREAU\nobreakdash-\hspace{0pt}2b, we can compare colors measured from our NIRC2 photometry to the intrinsic colors of M-type objects to estimate extinction at the time of our NIRC2 observations (assuming that the host star is not variable).

For KOINTREAU\nobreakdash-\hspace{0pt}1b, we calculate the expected color for a young M9$\pm$2 object by fitting a fifth-order polynomial to the spectral types and MKO $J-K$ colors of young objects in the UltracoolSheet, an online catalog of over 4000 ultracool dwarfs and imaged exoplanets, including photometry, parallaxes, and spectroscopic classifications from numerous sources. We find $(J-K)_0 = 1.23\pm0.23$ mag. By comparing this with our photometry, we infer $A_V=10\pm2$ mag, which is greater than but within $2\sigma$ of the 6.7 mag we find from spectral fitting.

For KOINTREAU\nobreakdash-\hspace{0pt}2b, we take the hybrid dwarf-giant spectra used in our spectral fitting process and synthesize their MKO $J-K$ colors. We adopt the M4.5 value for KOINTREAU\nobreakdash-\hspace{0pt}2b and use the spread of colors from M3.5--M6 to estimate the error on this value, giving $J-K=0.92\pm0.03$ mag. Comparing this with the observed $J-K$ photometry of KOINTREAU\nobreakdash-\hspace{0pt}2b indicates $A_V=6.8\pm0.7$ mag, which is $2.7\sigma$ inconsistent with the best-fit extinction $A_V=4.9$ mag from our spectral fitting, suggesting that extinction of this object is variable.

\subsection{Color-magnitude diagram}
\label{sec:discrepancy}

\begin{figure}[t]
    \centering
    \includegraphics[width=\linewidth]{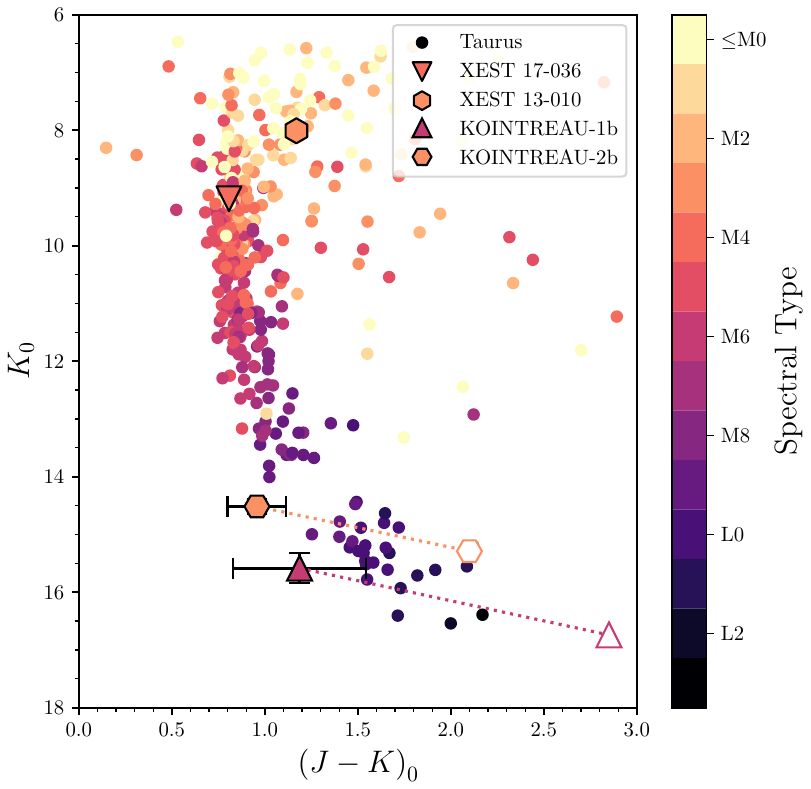}
    \caption{Extinction-corrected $J-K$ colour vs. extinction-corrected apparent $K$-band magnitude for our two candidate companions and their host stars, as well as Taurus stars with spectral types K4-L3 from \citet{esplin2019}. All objects are colored by their spectral type (Section~\ref{sec:spts}). The hollow markers indicate the unextincted position of each companion, and the dotted lines show their extinction vectors. We use extinction values derived from our photometry (Section~\ref{subsub:photextinction}) for our companions. All \citet{esplin2019} photometry from 2MASS has been transformed to the MKO system using relations given in \citet{skrutskie2006}.}
    \label{fig:tauruscmd_byspt}
\end{figure}

\begin{figure}[t]
    \centering
    \includegraphics[width=\linewidth]{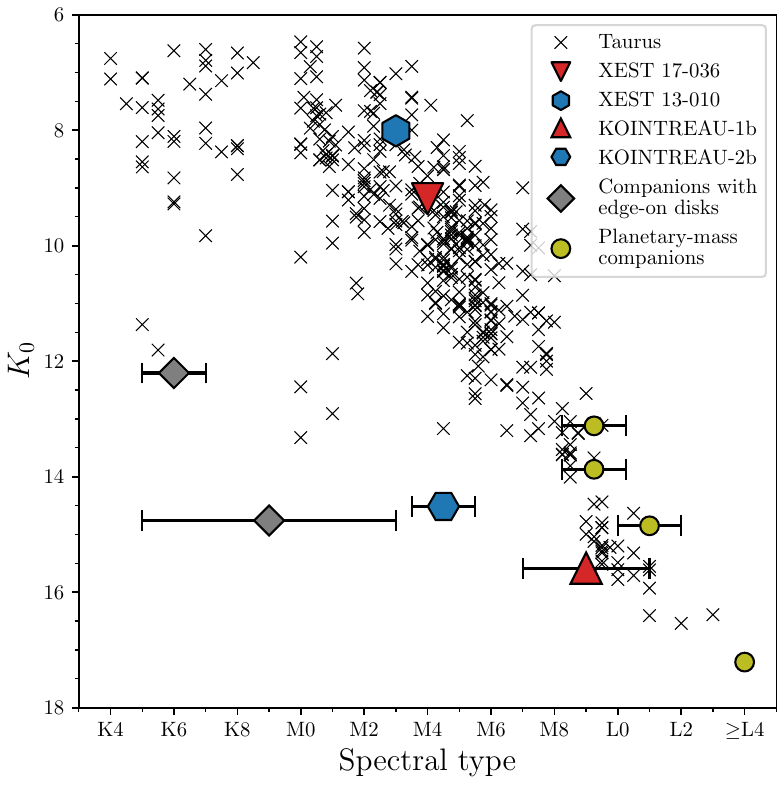}
    \caption{Extinction-corrected apparent $K$-band magnitude as a function of spectral type for our two companions and their host stars, as well as Taurus stars with spectral types K4-L3 from \citet{esplin2019}. Additional comparison objects are known companions in Taurus obscured by edge-on disks (from brightest to faintest: HK~Tau~B, \citeauthor{stapelfeldt1998}, \citeyear{stapelfeldt1998}; HV~Tau~C, \citeauthor{stapelfeldt2003}, \citeyear{stapelfeldt2003}; 2M~043726, \citeauthor{gaidos2022}, \citeyear{gaidos2022}) and planetary-mass companions (PMCs) in Taurus (from brightest to faintest: FU~Tau~B, \citeauthor{luhman2009futau}, \citeyear{luhman2009futau}; DH~Tau~B, \citeauthor{itoh2005}, \citeyear{itoh2005}; 2M~0441~Bb, \citeauthor{bowler2015aabbab}, \citeyear{bowler2015aabbab}; 2M~0437~b, \citeauthor{gaidos2022}, \citeyear{gaidos2022}).}
    \label{fig:tauruskmagvsspt}
\end{figure}

We couple our photometrically-derived extinctions with the apparent magnitudes from our photometry (Section~\ref{sec:imaging}) to place the companions on a dereddened color-magnitude diagram of Taurus sources (Figure~\ref{fig:tauruscmd_byspt}). We use our photometric extinctions to mitigate any impact from variable extinction, though adopting the spectroscopically-derived extinctions would not significantly impact our results. KOINTREAU\nobreakdash-\hspace{0pt}1b coincides with the few late-M/early L-type objects in the \citet{esplin2019} sample, and
KOINTREAU\nobreakdash-\hspace{0pt}2b sits in a region occupied by late-M Taurus objects.

KOINTREAU\nobreakdash-\hspace{0pt}2b's spectral type is notably discrepant with its color-magnitude diagram position. This is also apparent in Figure~\ref{fig:tauruskmagvsspt}, which shows that KOINTREAU\nobreakdash-\hspace{0pt}2b is significantly fainter than other Taurus sources of the same spectral type. The simplest explanation for this would be that the host star (and thus our companion) is incorrectly identified as a Taurus member. However, the \textit{Gaia} astrometry for XEST 13-010 has no flags associated with poor quality; the parallax agrees with the distances of Taurus members; and the proper motion agrees with its associated Taurus subcluster. We thus eliminate this possibility.

Another plausible reason is that the host star is comoving with but older than the main Taurus population. Ultracool objects cool over time, becoming fainter as they age \citep[e.g.,][]{sm08}. As such, younger objects of a given spectral type are more luminous than older ones.
Table \ref{tab:xest13} gives the ages of the L1529 subcluster that XEST 13-010 belongs to per \citet{kerr2021}, which at 3.4 Myr is slightly older than the 1-2 Myr age typically quoted for Taurus \citep{kenyon1995}. 
To test whether or not this age difference might explain the observed discrepancy between brightness and spectral type for KOINTREAU\nobreakdash-\hspace{0pt}2b, we linearly interpolate the BHAC15 model grid \citep{bhac2015} in log(age), log(effective temperature) and $K$-band magnitude. We find the effective temperature corresponding to a spectral type of M4.5 using Table 5 of \citet{herczeg2014}, and see how much the $K$-band magnitude varies between M4.5 objects that are 3.4 Myr and 1.5 Myr old. This approach yields an expected $K$-band magnitude difference of 0.8 magnitudes, which does not compensate for the $\sim$2.5 magnitude discrepancy we observe between KOINTREAU\nobreakdash-\hspace{0pt}2b and the faint end of the distribution of M4-M5 Taurus objects. In fact, we find that KOINTREAU\nobreakdash-\hspace{0pt}2b would have to be older than 500 Myr to simultaneously account for its spectral type and brightness, which is discordant with its low gravity spectrum. Therefore, KOINTREAU\nobreakdash-\hspace{0pt}2b must be experiencing some additional grey extinction in order to reconcile its observed properties with the age of Taurus (Section~\ref{sec:xest13interpretation}).

\subsection{Companion Masses}
\label{sec:masses}

Having determined that these objects are comoving companions to Taurus stars, we used the \textit{Gaia} DR3 parallaxes of the host stars to compute the companions' absolute magnitudes from their apparent magnitudes. These can be used to estimate the companions' masses using an evolutionary model. However, our analysis of KOINTREAU\nobreakdash-\hspace{0pt}2b indicates that this object has significant unaccounted-for grey extinction. Thus, we can only securely estimate the mass of KOINTREAU\nobreakdash-\hspace{0pt}1b in this manner.

We first computed KOINTREAU\nobreakdash-\hspace{0pt}1b's bolometric luminosity ($L_{\rm bol}$) using the MKO bolometric corrections for young ultracool objects as a function of spectral type reported in \citet{sanghi2023}. \citet{sanghi2023} caution that these corrections may not be valid for objects with ages $<10$Myr, but we use them given the absence of bolometric corrections for ultracool objects in star-forming regions like Taurus.

We linearly interpolated the \texttt{DUSTY} model grid \citep{dusty, dusty2} in log(age), log($L_{\rm bol}$) and log(mass). We created Gaussian distributions from the log of our computed $L_{\rm bol}$ and the log of our isochrone age for the host star given its Taurus subcluster membership \citep[$3.3\pm0.9$ Myr;][]{kerr2021}, using the associated errors as the standard deviation of each distribution. We then drew 10000 random samples from each distribution, which we used to obtain a posterior distribution of log(mass) values using \texttt{DUSTY}. We followed this process separately with the $L_{\rm bol}$ values obtained from the companion's $JHK$ magnitudes, then combined the posteriors from each magnitude into a single posterior for the mass of the companion. We took the median of this posterior to obtain the final mass value and the 68\% confidence interval as the uncertainty. From this process, we find that KOINTREAU\nobreakdash-\hspace{0pt}1b is $10.6^{+2.5}_{-2.3}$ M$_{\rm Jup}$. This value is below the canonical 13 M$_{\rm Jup}$ upper bound on exoplanet masses \citep{exoplanetdefn} and thus defines KOINTREAU\nobreakdash-\hspace{0pt}1b as a planetary-mass companion.

As additional validation, we repeated this process by comparing KOINTREAU\nobreakdash-\hspace{0pt}1b's observed absolute magnitudes with the model magnitudes from the \texttt{DUSTY} model grid, instead of computing KOINTREAU\nobreakdash-\hspace{0pt}1b's bolometric luminosity. The mass we find from this is $10.2^{+2.4}_{-2.1}$ M$_{\rm Jup}$, which is consistent with the value reported above. Thus, any offset from using the \citet{sanghi2023} bolometric corrections is well within our error bars.

\subsection{Emission lines in the spectrum of KOINTREAU\nobreakdash-\hspace{0pt}2b}
\label{sec:xest13he}

Both the Gemini/GNIRS and IRTF/SpeX spectra of KOINTREAU\nobreakdash-\hspace{0pt}2b feature a very prominent He~I 1.083 $\upmu$m emission line, which has a peak flux an order of magnitude greater than the continuum in the GNIRS spectrum. This line has a high excitation energy and thus its formation is restricted to regions of high temperature or regions near ionizing radiation, making it a useful tracer of infalling and outflowing gas in the inner regions of a circumstellar disk, although it may also arise from chromospheric emission \citep{edwards2006}. In particular, He~I found solely in emission (i.e. without any other component to its line profile) is taken to be an observational signature of jets or disk winds, seen in edge-on or sub-luminous accreting sources \citep{erkal2022}. This appears to be the case for the line we observe in KOINTREAU\nobreakdash-\hspace{0pt}2b, although higher spectral resolution is required to fully resolve the line and confirm its profile as our data has a maximum velocity resolution of $\sim$170kms$^{-1}$, which is higher than the $\lesssim$100kms$^{-1}$ centroid velocities of emission-only profiles shown in Figure 12 of \citet{erkal2022}. 

\citet{alcala2014} finds a correlation between the accretion luminosity and He~I line luminosity of young stars in Lupus, a star-forming region of similar age to Taurus \citep{galli2020}. We use this relation to calculate the accretion luminosity of KOINTREAU\nobreakdash-\hspace{0pt}2b, assuming the observed He~I emission does indeed arise from ongoing accretion. To do this, we flux-calibrate our Gemini/GNIRS spectrum using our Keck/NIRC2 $J$-band photometry, deredden the spectrum using the best-fit extinction from our spectral fitting, and integrate the He~I line flux, obtaining $3.165 \pm 0.005 \times 10^{-17}$ Wm$^{-2}$. Using the \textit{Gaia} distance to the host star, we convert this to a line luminosity and then to an accretion luminosity given the \citet{alcala2014} relation between He~I 1.083 $\upmu$m line luminosity and accretion luminosity. This gives $\log\left({L}_{\rm acc}/{L}_{\odot}\right)=-2.7\pm0.4$ dex, where the error on this measurement is a quadrature sum of a Monte Carlo error estimate for the line luminosity and the scatter on the \citet{alcala2014} accretion luminosity - line luminosity relation.

\citet{alcala2014} also present a relation between accretion luminosity and mass accretion rate in their Equation 1. Essentially an expression of conservation of energy from the infall of accreting material, this relation gives that
\begin{equation}
    \dot{M}_{\text{acc}} = \left( 1 - \frac{R_{\star}}{R_{\text{in}}} \right)^{-1} \frac{L_{\text{acc}} R_{\star}}{G M_{\star}} \approx 1.25 \frac{L_{\text{acc}} R_{\star}}{G M_{\star}},
\end{equation}
where \citet{alcala2014} derive the right-hand side of the equation by adopting a standard value of infall radius $R_{\text{in}} = 5R_{\star}$ for consistency with previous literature results.

Using this equation to estimate KOINTREAU\nobreakdash-\hspace{0pt}2b's mass accretion rate requires knowledge of its mass and radius. KOINTREAU\nobreakdash-\hspace{0pt}2b's subluminous nature precludes us from computing these values directly from our photometry, but we can use its spectral type and age to estimate its mass and radius by linearly interpolating the BHAC15 evolutionary model grid in log(age), log($T_{\rm eff}$) and log(mass). This gives a mass of $0.35\pm0.04$ ${\rm M}_\odot$ and a radius of $1.08\pm0.07$ ${\rm R}_\odot$. Using these values, we estimate a mass accretion rate of $\log\left({\rm M}_{\rm acc} / {\rm M}_\odot{\rm yr}^{-1} \right)= -9.6 \pm 0.4$ dex. This falls below the value estimated for an object of this mass by both the \citet{hh2008} and \citet{gangi2022} accretion rate-stellar mass relations derived for Taurus objects by 2.2$\sigma$ and 0.9$\sigma$, respectively. This could imply at least one of four things: that its mass accretion rate is inherently lower than other objects, that the line flux is obscured by the discussed grey extinction despite its overwhelmingly large flux compared to the continuum, that the mass of KOINTREAU\nobreakdash-\hspace{0pt}2b is not what is expected given its spectral type, or that the emission line is indeed chromospheric in origin.

In addition to the He~I 1.08 $\upmu$m line, there are several H$_2$ lines evident in the $K$-band region of the spectrum (Figure~\ref{fig:xest13spec}). These emission lines have their origins either in thermal excitation from shocks \citep{greene2010} or fluorescence induced by far-ultraviolet (FUV) emission from the host star \citep{doyon1994, burton2002}. The former of these two cases would agree with the accretion shock origin of the He~I line.

\begin{figure}
    \centering
    \includegraphics[keepaspectratio,width=\linewidth]{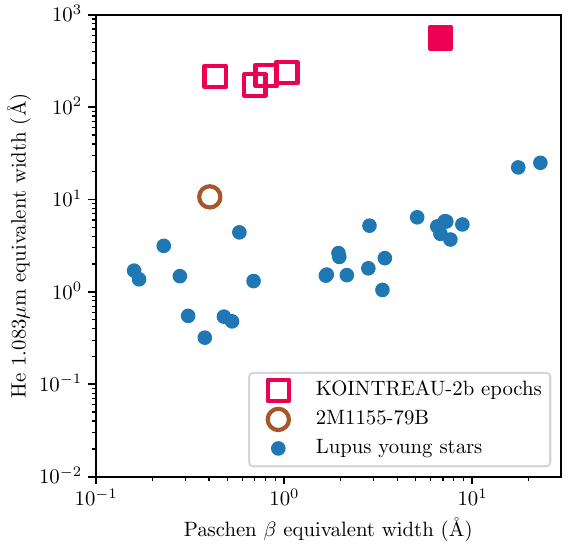}
    \caption{He 1.083 $\upmu$m line equivalent width against Paschen $\beta$ equivalent width for each epoch of spectra of KOINTREAU\nobreakdash-\hspace{0pt}2b, as well as the similar object 2M1155-79B \citep{dicksonvdv2022}, compared to young stars in Lupus from \citet{alcala2014}. KOINTREAU\nobreakdash-\hspace{0pt}2b consistently has an He~I line equivalent width that is an order of magnitude larger than any of these other objects. KOINTREAU\nobreakdash-\hspace{0pt}2b and 2M1155-79B values were measured for this paper. For these, filled markers indicate Paschen $\beta$ detections, and hollow markers indicate Paschen $\beta$ non-detections, judged by inspection. Pa$\beta$ values for hollow markers are 3$\sigma$ upper limits.}
    \label{fig:heews}
\end{figure}

It is curious to note that hydrogen emission lines are typically a hallmark of accretion, but we do not see these in KOINTREAU\nobreakdash-\hspace{0pt}2b's spectrum. To investigate how peculiar this is, we compare the relationship between the equivalent width of the He~I 1.083~$\upmu$m line and the equivalent width of the Paschen $\beta$ (Pa$\beta$) line for the five individual epochs of spectra for KOINTREAU\nobreakdash-\hspace{0pt}2b to the young Lupus stars from \citet{alcala2014} in Figure~\ref{fig:heews}. We can see that the He 1.083 $\upmu$m equivalent width of KOINTREAU\nobreakdash-\hspace{0pt}2b is anomalously large compared to the \citet{alcala2014} sample. Additionally, in the \citet{alcala2014} sample He~I and Pa$\beta$ equivalent widths are correlated, whereas we observe huge He~I emission without the expected corresponding Pa$\beta$ emission, with the exception of the UT 2024 January 23 epoch. The only epoch of KOINTREAU\nobreakdash-\hspace{0pt}2b data with strong Pa$\beta$ emission is also the epoch with the strongest He~I emission, which may imply that He~I and Pa$\beta$ are still correlated for our object but that there is some physical mechanism suppressing the Pa$\beta$ flux relative to the He~I flux.
We discuss the possible physical scenarios that might cause this anomalous He/Pa$\beta$ equivalent-width ratio in Section~\ref{sec:xest13interpretation}.

\section{Physical Interpretation}

\subsection{KOINTREAU\nobreakdash-\hspace{0pt}1b}
\label{sec:xest17interpretation}
\begin{figure}
    \centering
    \includegraphics[width=\linewidth]{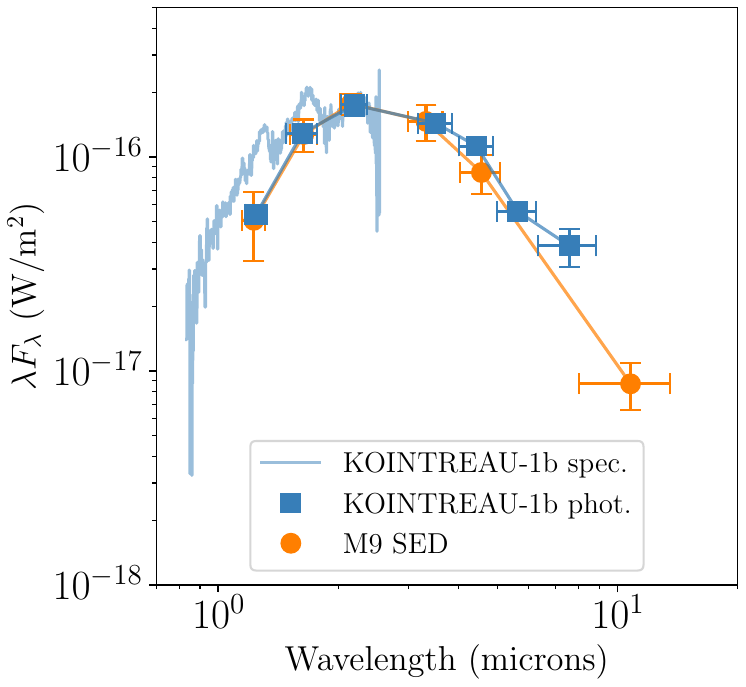}
    \caption{Spectral energy distribution (SED) of KOINTREAU\nobreakdash-\hspace{0pt}1b compared to an SED for a young M9 dwarf, constructed using the \citet{best2018} color-SpT relations, calibrated using the dereddened $K$-band magnitude from our photometry and then reddened using our photometrically-derived $A_V$ (Section~\ref{subsub:photextinction}). Our IRTF/SpeX spectrum is scaled to KOINTREAU\nobreakdash-\hspace{0pt}1b's observed $K$-band flux. The $x$-axis error bars on each photometry point indicate the width of the corresponding bandpass. The difference in slope between KOINTREAU\nobreakdash-\hspace{0pt}1b's spectrum and photometry reflects the differing extinction between the two epochs (Section \ref{subsub:photextinction}).}
    \label{fig:xest17sed}
\end{figure}

KOINTREAU\nobreakdash-\hspace{0pt}1b appears from our spectra to be a typical young planetary-mass companion, with no prominent He or H emission. However, the differing slopes between the Gemini/GNIRS and IRTF SpeX/Prism spectra, as well as our variable $K$-band photometry, suggest that KOINTREAU\nobreakdash-\hspace{0pt}1b either experiences variable levels of extinction or is intrinsically variable. 
We use the totality of our data for KOINTREAU\nobreakdash-\hspace{0pt}1b to construct its spectral energy distribution (SED, Figure~\ref{fig:xest17sed}). We compare this with an SED for a young M9 dwarf from \citet{best2018}, normalized using the dereddened $K$-band magnitude from our photometry and then reddened using the photometric extinction. We note that since the \citet{cardelli1989} extinction law is only valid for wavelengths 0.3 $\upmu\rm{m} < \lambda < $ 3.3 $ \upmu\rm{m}$, we use the Wide-field Infrared Survey Explorer
(WISE) W1, W2, and W3 $A_{\lambda}/A_V$ values presented in \citet{wang2019} for these bands. Figure~\ref{fig:xest17sed} shows that the SED of KOINTREAU\nobreakdash-\hspace{0pt}1b is brighter at its longest wavelengths than the empirical SED -- this, coupled with the observed variable spectral slope (Figure~\ref{fig:xest17speccomp}), suggest the presence of a disk around the companion, but we would require longer-wavelength resolved photometry of the XEST 17-036 system to be able to draw definitive conclusions.
This variability might also be intrinsic, as substellar objects are known to be variable on timescales as short as a few hours, which is generally attributed to their cloudy atmospheres \citep[e.g.][]{artigau2018}. This would readily explain KOINTREAU\nobreakdash-\hspace{0pt}1b's changing brightness, and spectral variability has been seen to vary as a function of wavelength in later-type brown dwarfs \citep[e.g.,][]{lew2016, milespaez2019}, although it is unclear if this could induce the observed change in spectral slope.

KOINTREAU\nobreakdash-\hspace{0pt}1b is the fifth planetary-mass companion to be found in Taurus. Figure~\ref{fig:tauruskmagvsspt} shows that KOINTREAU\nobreakdash-\hspace{0pt}1b has a similar spectral type to three of these (DH Tau B, FU Tau B, and 2M 0441 Bb) and is fainter than all three, making it likely the lowest-mass companion of its spectral type found in Taurus.

\subsection{KOINTREAU\nobreakdash-\hspace{0pt}2b}
\label{sec:xest13interpretation}

We consider three physical scenarios that might cause the apparent underluminosity of KOINTREAU\nobreakdash-\hspace{0pt}2b relative to other objects of the same spectral type in Taurus (Section~\ref{sec:discrepancy}).
The first of these, that the system is older (and thus fainter) than the average Taurus early-M dwarf, is already disfavored as KOINTREAU\nobreakdash-\hspace{0pt}2b would have to be older than 500 Myr to simultaneously account for its spectral type and brightness given the measured extinction, which does not agree with its low-gravity spectrum (Section~\ref{sec:discrepancy}). 
The second hypothesis is that KOINTREAU\nobreakdash-\hspace{0pt}2b is a highly embedded Class I source, as the shrouding of this object by its natal disk and envelope would provide a natural explanation for its underluminosity. However, if this were the case we would expect to see a slope in our NIR spectra commensurate with $A_V\approx30$-$60$ mag \citep{greene1996}. We do not observe this, and thus we rule out the possibility.

We therefore adopt the remaining plausible physical scenario given our data, which is that KOINTREAU\nobreakdash-\hspace{0pt}2b is obscured by a highly inclined (edge-on) disk and thus seen mostly in scattered light. The $y-J$ vs. $J-K$ color-color diagram position of KOINTREAU\nobreakdash-\hspace{0pt}2b agrees with that of other known edge-on disks in Taurus \citep[][Fig. 38]{zhang2018}, lending additional support to this conclusion.

Given this physical picture, there are three cases that could cause the anomalous He/H emission ratio discussed in Section~\ref{sec:xest13he}. The first is that the He~I emission is seen directly from an extended emission jet above/below the edge-on disk, whereas the continuum and atomic hydrogen emission are seen in reflected light -- this has caused large emission-line equivalent widths in other known edge-on disks \citep[e.g., HH 30;][]{wood1998} and could feasibly cause the emission ratio we observe, although the magnitude of this effect is such that the H equivalent width would still have to be intrinsically small. There is also the second possibility that the He and H emission are both seen either in reflected light or directly. This would mean that the anomalous ratio between the two is intrinsic, perhaps connected to the ultraviolet flux of KOINTREAU\nobreakdash-\hspace{0pt}2b itself, but explaining this would require a plasma physics explanation that is beyond the scope of this work. We note that there are no [Fe II] lines in our spectra of KOINTREAU\nobreakdash-\hspace{0pt}2b, which would provide key support for an extended jet that might produce the He~I emission, nor do our NIRC2 images show evidence of extended emission (although deeper narrow-band imaging would be needed to rule this out).

With our edge-on diagnosis, KOINTREAU\nobreakdash-\hspace{0pt}2b joins a small number of subluminous young objects attributed to obscuration by an edge-on disk \citep[]{stapelfeldt1998, stapelfeldt2003, luhman2004, 2007mohanty, looper2010, zhang2018, haffert2020, christiaens2021, flores2021, dicksonvdv2022, gaidos2022}. All of these objects are several magnitudes fainter than their spectral types, ages and distances would suggest. The NIR spectra of edge-on disks Oph~163131 \citep{flores2021}, TWA~30B \citep{looper2010, venuti2019}, and 2M~1155\nobreakdash-\hspace{0pt}79B \citep{dicksonvdv2022} all show strong He~I 1.08 $\upmu$m emission and lack significant H emission, as with KOINTREAU\nobreakdash-\hspace{0pt}2b (Figure~\ref{fig:heews}). The 1.08 $\upmu$m emission detected in CrA\nobreakdash-\hspace{0pt}9~b \citep{christiaens2021}, another of these edge-on systems, is tentatively attributed by the authors to Paschen $\gamma$, but we note that the $R\sim50$ resolution of the VLT/SPHERE IFS prevents distinguishing if the emission is from He~I or Paschen $\gamma$, as both these lines fall within the same resolution element. 
Further, the edge-on system 2MASS~J04372631+2651438 \citep{gaidos2022} shows little-to-no evidence for emission at 1.08 $\upmu$m, nor indeed anywhere else in the 0.8-2.4 $\upmu$m spectrum.

A number of the edge-on disks referenced above are within Taurus, namely HK~Tau~B \citep{stapelfeldt1998}, HV~Tau~C \citep{stapelfeldt2003}, 2MASS~J04185813+2812234, GM~Tau, ITG~3 \citep{zhang2018}, and 2MASS~J04372631+2651438 \citep{gaidos2022}. Of these, HK~Tau~B, HV~Tau~C, and 2MASS~J04372631+2651438 are all companions to brighter stars, as with KOINTREAU\nobreakdash-\hspace{0pt}2b. HK Tau B and HV Tau C have both had their disks resolved by HST imaging, and both also have $K$- and $L$-band spectra that show a deep water ice absorption feature at 3.1 $\upmu$m \citep{terada2007}. In the case of HV Tau C, the $K$-band spectrum also shows several of the same H$_2$ emission lines that we see in KOINTREAU\nobreakdash-\hspace{0pt}2b, which are attributed to jet excitation from strong outflows from the object \citep{magazzu1994, woitas1998, terada2007}, in agreement with our physical picture of KOINTREAU\nobreakdash-\hspace{0pt}2b. As already mentioned, the IRTF/SpeX Prism spectrum of 2MASS~J04372631+2651438 does not feature any prominent emission lines, implying that it may not be accreting as strongly as some of the other systems we have discussed -- although \citet{gaidos2022} note that this object has elevated Pan-STARRS $r$-band flux compared to template SEDs, which could be attributed to H$\alpha$ accretion.

\begin{figure}
    \centering
    \includegraphics[width=\linewidth]{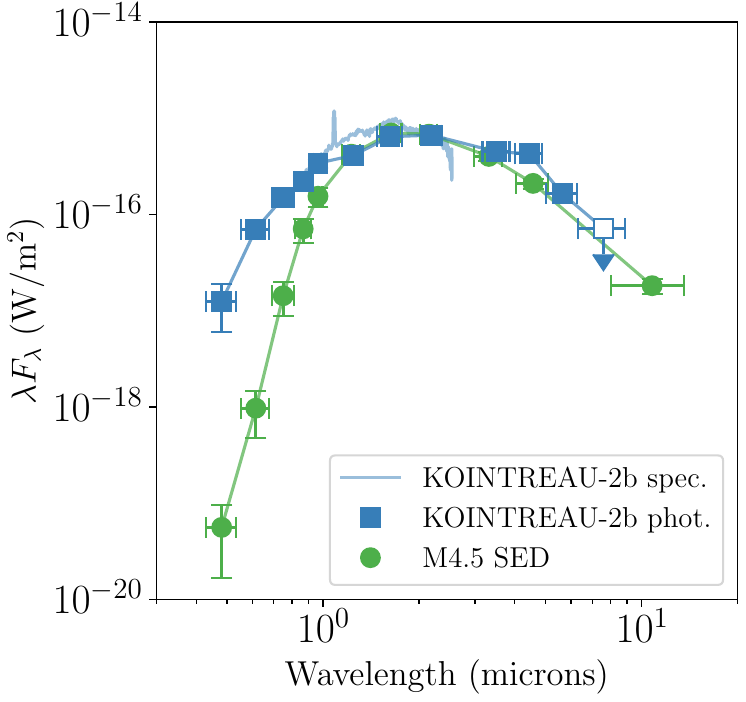}
    \caption{Spectral energy distribution (SED) of KOINTREAU\nobreakdash-\hspace{0pt}2b compared to an SED formed by averaging the M4 and M5 colors reported in \citet{best2018}, calibrated using the dereddened $K$-band magnitude from our photometry and then reddened using our photometrically-derived $A_V$ (Section~\ref{subsub:photextinction}). Our IRTF/SpeX spectrum is scaled to KOINTREAU\nobreakdash-\hspace{0pt}2b's observed $K$-band flux. The $x$-axis error bars on each photometry point indicate the width of the corresponding bandpass.}
    \label{fig:xest13sed}
\end{figure}

 Figure~\ref{fig:xest13sed} compares the SED of KOINTREAU\nobreakdash-\hspace{0pt}2b to the average of M4 and M5 SEDs constructed using the \citet{best2018} color-SpT relations and calibrated using the $K$-band magnitude from our photometry as with KOINTREAU\nobreakdash-\hspace{0pt}2b.
Figure~\ref{fig:xest13sed} shows that our NIRC2 photometry is well-matched by the empirical SED. At longer wavelengths, we see an excess relative to the empirical M4.5 SED in \textit{Spitzer}/IRAC [4.5] (corresponding to a peak blackbody temperature $\sim$650K), but this does not continue in [5.8] (and [8.0] has no detection). However, the flux from an edge-on disk may only become prominent at wavelengths longer than [8.0] \citep[e.g.,][]{duchene2010, dicksonvdv2022}. Thus KOINTREAU\nobreakdash-\hspace{0pt}2b's non-detection in this band does not preclude it from having a disk. Our optical photometry is brighter than predicted by our comparison SED, which we attribute to scattering of KOINTREAU\nobreakdash-\hspace{0pt}2b's light by its disk, although it is possible that H$\alpha$ emission is also a contributing factor to the $r$-band as for 2MASS~J04372631+2651438. 

To properly constrain the properties of such an edge-on disk, we would require mid-infrared photometry ($>$10~$\upmu$m). However, given that KOINTREAU\nobreakdash-\hspace{0pt}2b is only $4.3^{\prime\prime}$ from its host star (which has its own disk), existing observations of this system in WISE bands W3 and W4 \citep{wright2010} and \textit{Spitzer}'s Multiband Imaging Photometer (MIPS) 24 $\upmu$m band \citep{padgett2004prop} are insufficient to resolve the two sources. 
\citet{dicksonvdv2022} report a centroid shift of $4.14\arcsec$ in WISE photometry from the host star to 2M1155-7919B between bands W1 to W4, which they use to argue that the W4 flux comes primarily from the companion, which provides them with long-wavelength photometry for its disk. However, we do not see any comparable centroid shift in the WISE data for KOINTREAU\nobreakdash-\hspace{0pt}2b, presumably because XEST 13-010's bright disk mitigates the W4 centroid shift induced by KOINTREAU\nobreakdash-\hspace{0pt}2b's disk.
The only additional constraints on the disk of KOINTREAU\nobreakdash-\hspace{0pt}2b come from \citet{andrews2013}, who observed the XEST 13-010 system using the Submillimeter Array in its compact configuration, but did not note a binary component. Thus, we assume \citet{andrews2013} did not detect KOINTREAU\nobreakdash-\hspace{0pt}2b at the 0.2 mJy/beam rms. This implies a 3$\sigma$ upper limit on the dust mass of KOINTREAU\nobreakdash-\hspace{0pt}2b's disk of 1.2 M$_{\rm Earth}$, which is scaled from the \citet{akeson2019} value for the primary star's disk dust mass. This value is on the lower end but within the spread of dust masses that \citet{wardduong2018} found for Taurus stars of KOINTREAU\nobreakdash-\hspace{0pt}2b's estimated mass (see their Figure~11).


\section{Conclusion}

We have discovered two new companions to young stars XEST 17-036 and XEST 13-010 in Taurus. By imaging these systems over multiple years using Keck/NIRC2 with adaptive optics, we measured the movement of each companion relative to its host star, and when combined with \textit{Gaia} astrometry for the host stars we establish that both companions are gravitationally bound. We obtained Keck/NIRC2 $JHK$ photometry, as well as IRTF/SpeX Prism ($R\sim100$) and Gemini/GNIRS ($R\sim1000$-2000) spectra, for both KOINTREAU\nobreakdash-\hspace{0pt}1b (the companion to XEST 17-036) and KOINTREAU\nobreakdash-\hspace{0pt}2b (the companion to XEST 13-010). We also extracted Pan-STARRS and \textit{Spitzer}/IRAC photometry for these companions where able. 

We used these data to determine spectral types and $A_V$ for both objects by comparison with literature spectra. This gave a spectral type of M9 $\pm$ 2 for KOINTREAU\nobreakdash-\hspace{0pt}1b and a spectral type of M4.5 $\pm$ 1 for KOINTREAU\nobreakdash-\hspace{0pt}2b. 

Using the \texttt{DUSTY} evolutionary model grids, we derived an estimated mass of $10.6^{+2.5}_{-2.3}$ M$_{\rm Jup}$ for KOINTREAU\nobreakdash-\hspace{0pt}1b, making it the fifth planetary-mass companion discovered in Taurus. The slope of KOINTREAU\nobreakdash-\hspace{0pt}1b's spectrum varies between epochs, which may indicate either atmospheric clouds or else a disk around this object. 

KOINTREAU\nobreakdash-\hspace{0pt}2b is the faintest object of its spectral type in Taurus by $>$2.5 magnitudes, which points to its nature as a young star occulted by an edge-on disk. Additionally, KOINTREAU\nobreakdash-\hspace{0pt}2b exhibits an exceptionally strong He~I 1.08 $\upmu$m emission line in both its IRTF/SpeX Prism and Gemini/GNIRS spectra compared to accreting young stars in Lupus \citep{alcala2014}, which show that strong He~I emission is typically accompanied by H emission. H emission is not present in KOINTREAU\nobreakdash-\hspace{0pt}2b's spectra, making it an unusual system.

These two discoveries add to the populations of planetary-mass and disk-obscured companions in Taurus. As the KOINTREAU survey continues we hope to find more planetary-mass companions in both Taurus and Ophiuchus, providing valuable anchors for the earliest stages in the evolution of substellar objects and helping to expand the study of extremely young directly imaged companions.

\begin{acknowledgments}

The authors wish to thank Eugene Magnier for helpful discussion regarding the use of Pan-STARRS data.

This research was funded in part by the Gordon and Betty Moore Foundation through grant GBMF8550 to M.~Liu.

Some of the data presented herein were obtained at Keck Observatory, which is a private 501(c)3 non-profit organization operated as a scientific partnership among the California Institute of Technology, the University of California, and the National Aeronautics and Space Administration. The Observatory was made possible by the generous financial support of the W. M. Keck Foundation.

This paper is partly based on observations obtained at Gemini Observatory, a program of NSF NOIRLab, which is managed by the Association of Universities for Research in Astronomy (AURA) under a cooperative agreement with the U.S. National Science Foundation on behalf of the Gemini Observatory partnership: the U.S. National Science Foundation (United States), National Research Council (Canada), Agencia Nacional de Investigaci\'{o}n y Desarrollo (Chile), Ministerio de Ciencia, Tecnolog\'{i}a e Innovaci\'{o}n (Argentina), Minist\'{e}rio da Ci\^{e}ncia, Tecnologia, Inova\c{c}\~{o}es e Comunica\c{c}\~{o}es (Brazil), and Korea Astronomy and Space Science Institute (Republic of Korea).

This paper has made use of the SpeX instrument \citep{2003PASP..115..362R} at the Infrared Telescope Facility, which is operated by the University of Hawaii under contract 80HQTR24DA010.

This work has made use of the Pan-STARRS1 Surveys (PS1) and the PS1 public science archive.

This work is based in part on observations made with the Spitzer Space Telescope, which was operated by JPL under a contract with NASA.

This work has made use of data from the European Space Agency (ESA) mission \href{https://www.cosmos.esa.int/gaia}{\it Gaia}, processed by the {\it Gaia} Data Processing and Analysis Consortium (\href{https://www.cosmos.esa.int/web/gaia/dpac/consortium}{DPAC}). Funding for the DPAC has been provided by national institutions, in particular the institutions participating in the {\it Gaia} Multilateral Agreement.

This work has benefited from \href{http://bit.ly/UltracoolSheet}{The UltracoolSheet}, maintained by Will Best, Trent Dupuy, Michael Liu, Aniket Sanghi, Rob Siverd, and Zhoujian Zhang, and developed from compilations by \citet{dupuy2012}, \citet{dupuy2013}, \citet{deacon2014}, \citet{liu2016}, \citet{best2018}, \citet{best2021}, \citet{sanghi2023}, and \citet{schneider2023}.

This work made use of Astropy \citep{astropy1, astropy2, astropy3}, and \texttt{ccdproc}, an Astropy package for image reduction \citep{ccdproc}.

Finally, the authors wish to recognize and acknowledge the very significant cultural role and reverence that the summit of Maunakea has always had within the Native Hawaiian community. We are grateful for the privilege of observing the Universe from a place that is unique in both its astronomical quality and its cultural significance.
\end{acknowledgments}

%

\facilities{Keck (NIRC2), IRTF (SpeX), Gemini (GNIRS), Pan-STARRS, \textit{Spitzer} (IRAC)}


\software{astropy \citep{astropy1, astropy2, astropy3}; SpectRes \citep{spectres}; ccdproc \citep{ccdproc}; SpeXtool \citep{2004PASP..116..362C}; PypeIt \citep{bochanski2009, bernstein2015, prochaska2020a, prochaska2020b}
}

\FloatBarrier



\newpage
\appendix
\section{NIRC2 Optical Distortion with the Pyramid Wavefront Sensor}
\label{ap:distortion}

The Keck/NIRC2 system has a small (sub-pixel) amount of known optical distortion, mostly consisting of higher-order variation. P. Brian Cameron created a NIRC2 distortion solution with data obtained using a pinhole mask circa 2006\footnote{\url{www2.keck.hawaii.edu/inst/nirc2/astrometry/distortion.pdf}}, and the first distortion solution using on-sky data was created by \citet{yelda2010}, who compared NIRC2 images of M92 to distortion-corrected HST ACS/WFC data. After NIRC2 instrument servicing in 2015, a revised distortion solution was produced by \citet{service2016} by comparing NIRC2 data of M53 with HST ACS/WFC3 data, and this solution has been in use ever since. However, there has been no published measurement of NIRC2's optical distortion when using the Pyramid wavefront sensor (PyWFS), which was installed in 2018 \citep{bond2018} and decommissioned in 2024. The optical paths when observing with the regular Shack-Hartmann wavefront sensor (SHWFS) and the PyWFS differ slightly due to the introduction of a dichroic that sends $J$- and $H$-band light to the PyWFS and $K$-band light to the NIRC2 detector. Therefore, it is possible that a change in optical distortion from the existing \citeauthor{service2016} solution might occur when using the PyWFS, which would then affect our astrometric analysis.

In order to understand the effect of switching between the Shack-Hartmann and Pyramid wavefront sensors, we compared Keck/NIRC2 narrow camera PyWFS data of M92 that we took between 2019-2022 to data collected of the same field by \citet{yelda2010} for the construction of their NIRC2 distortion solution.
Our motivation for this direct comparison between NIRC2 datasets is that it is simpler and allows for easier comparison than rederiving an astrometric solution by comparing our PyWFS images to HST data, as \citet{yelda2010} and \citet{service2016} did in the construction of their distortion solutions.

\begin{figure*}
\gridline{\fig{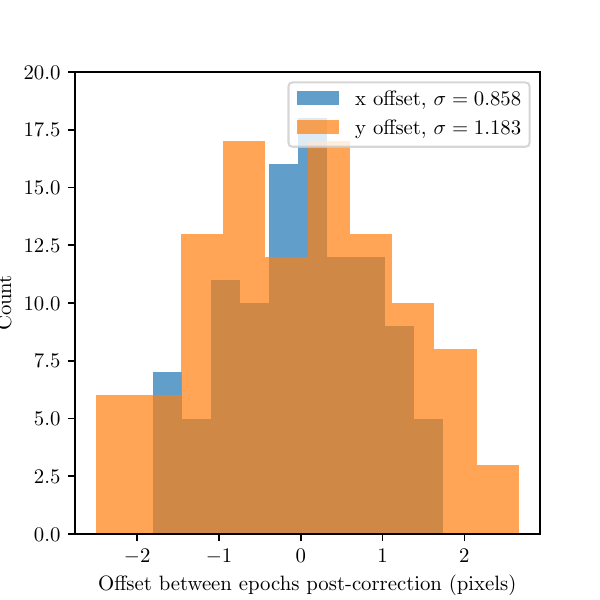}{0.45\linewidth}{(a)}
          \fig{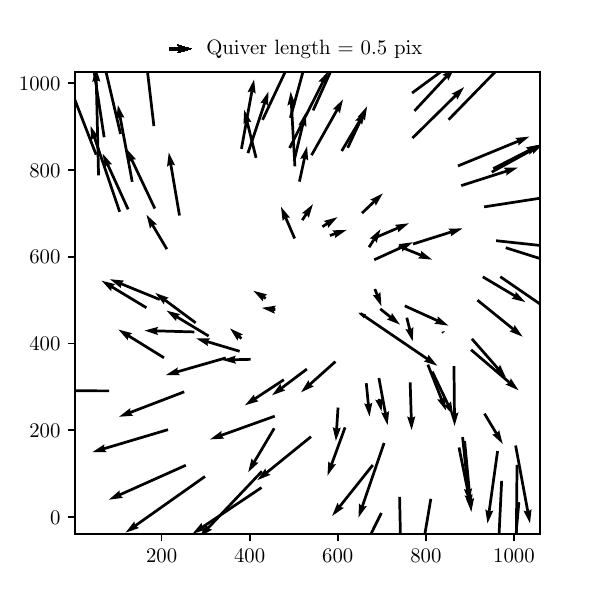}{0.45\linewidth}{(b)}}
\gridline{\fig{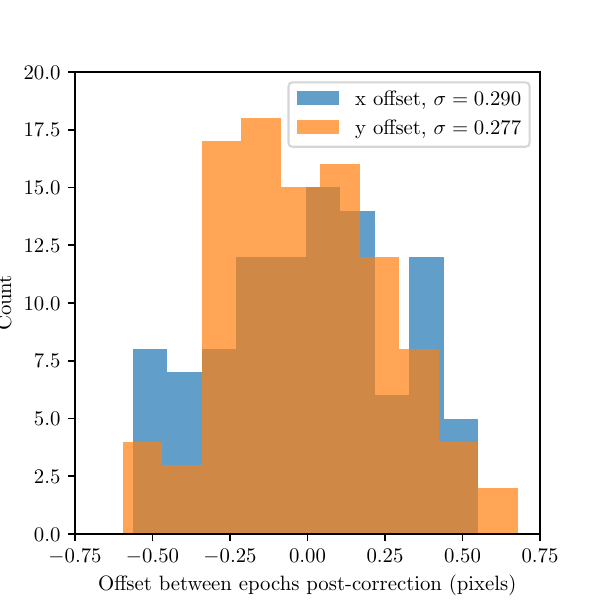}{0.45\linewidth}{(c)}
          \fig{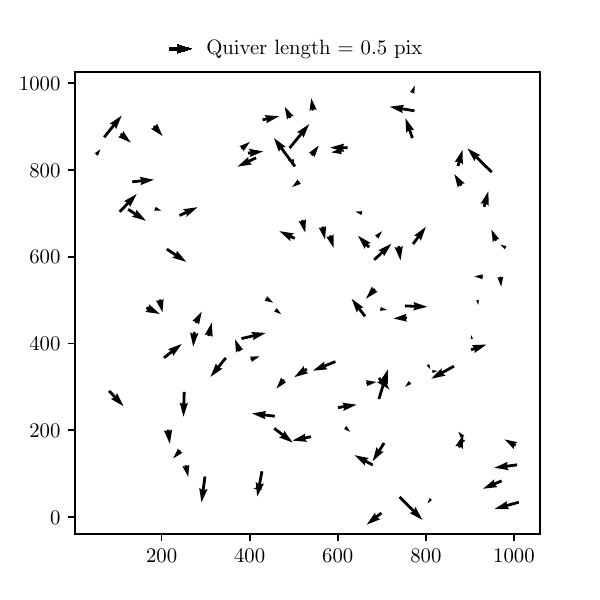}{0.45\linewidth}{(d)}}
\caption{Histogram of $xy$ offsets and quiver plot showing the difference between distortion-corrected 
pixel positions of stars common to the \citet{yelda2010} and 2020 PyWFS datasets. Figures (a) and (b) show the residuals when only correcting for an $xy$ shift between the two datasets, whereas Figures (c) and (d) show the residuals when correcting for an $xy$ shift, rotation and scale change. It is evident that the more complex correction describes the differences between the two datasets significantly better from the decreased scatter in the fit residuals. Note the larger $x$-axis in Figure (a) compared to Figure (c).}
\label{fig:yeldato2020}
\end{figure*}

To compare the two datasets, we retrieved the \citeauthor{yelda2010} data from the Keck Observatory Archive and then constructed matched starlists consisting of the $(x,y)$ coordinates of $106$ stars well-detected in both the \citet{yelda2010} data and our highest-quality PyWFS dataset from May 2020.
We corrected each starlist with the best available distortion corrections (i.e., the \citeauthor{yelda2010} distortion correction for the \citealp{yelda2010} data and the \citeauthor{service2016} correction for the PyWFS data) before comparing them. We initially performed a straight-line fit to determine the $x$- and $y$-offsets between the pixel coordinates of two datasets (2 free parameters), but found that this was insufficient to describe the transformation from the \citet{yelda2010} data to our PyWFS data (Figure \ref{fig:yeldato2020}). Therefore, we used an algorithm developed for matching molecules (the \href{https://github.com/jewettaij/superpose3d}{\texttt{superpose3d}} Python code based on the work of \citealp{Diamond1988}) to determine the $xy$ translation, rotation, and enlargement scale factor (4 free parameters) that best transformed one dataset to another. With both fitting methods we performed a round of sigma-clipping, removing from the fitting dataset any stars whose positions are $>2$ mean absolute deviations from the median in either the $x$ or $y$ directions -- this removed one outlier in the straight-line (2-parameter) fit case, and seven outliers when performing the full 4-parameter fit. In order to estimate errors on the fitted parameters, we bootstrap resampled with replacement our set of stars 10,000 times, and adopted the standard deviation of the resulting distributions for each parameter as its error.

\begin{figure*}
\gridline{\fig{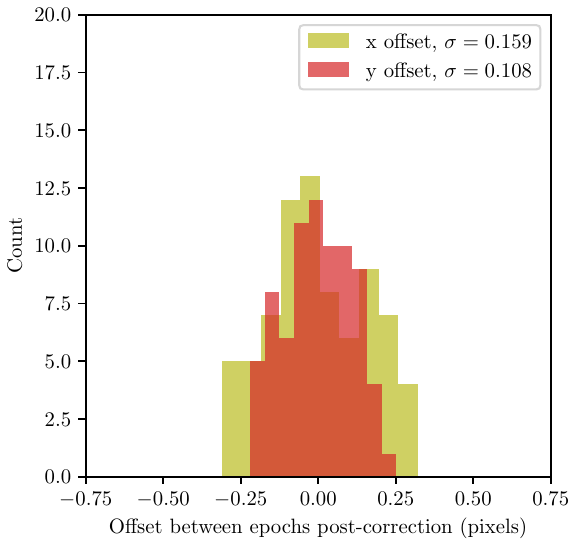}{0.45\linewidth}{(a)}
          \fig{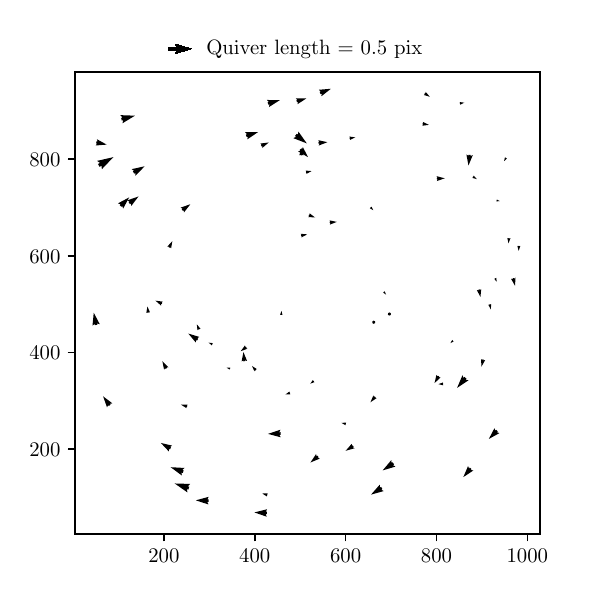}{0.45\linewidth}{(b)}}
\gridline{\fig{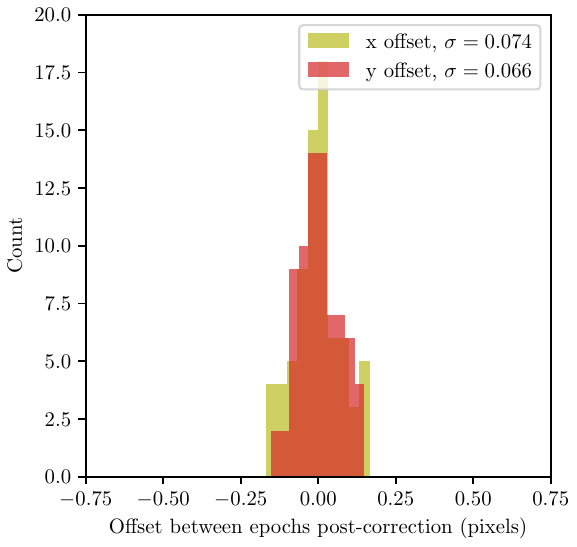}{0.45\linewidth}{(c)}
          \fig{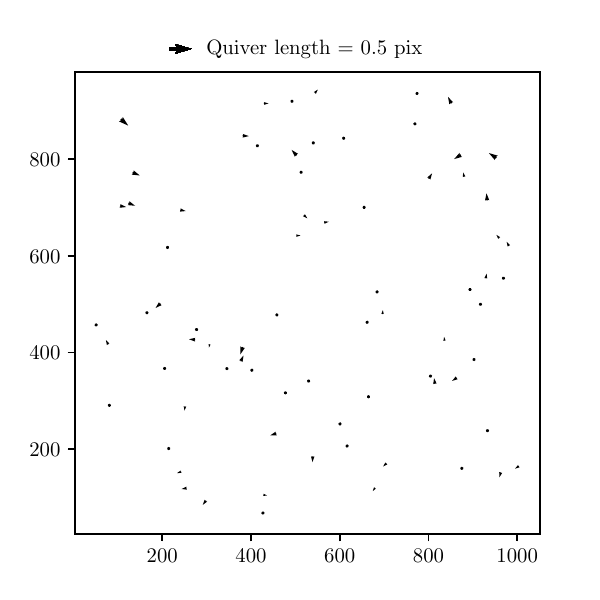}{0.45\linewidth}{(d)}}
\caption{Histogram of $xy$ offsets and quiver plot showing the difference between distortion-corrected 
pixel positions of stars common to the 2019 and 2020 PyWFS datasets. Figures (a) and (b) show the residuals when only correcting for an $xy$ shift between the two datasets, whereas Figures (c) and (d) show the residuals when correcting for an $xy$ shift, rotation and enlargement. The $0.025^\circ \pm 0.002^\circ$ rotational offset between the datasets is evident in (b) and sufficiently corrected in (c) and (d), as evidenced by the decreased scatter of the residuals.}
\label{fig:2019vs2020}
\end{figure*}
\begin{figure*}
\gridline{\fig{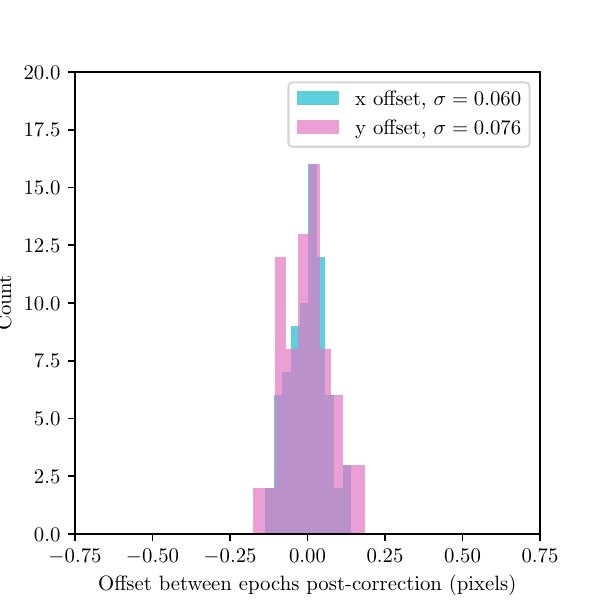}{0.45\linewidth}{}
          \fig{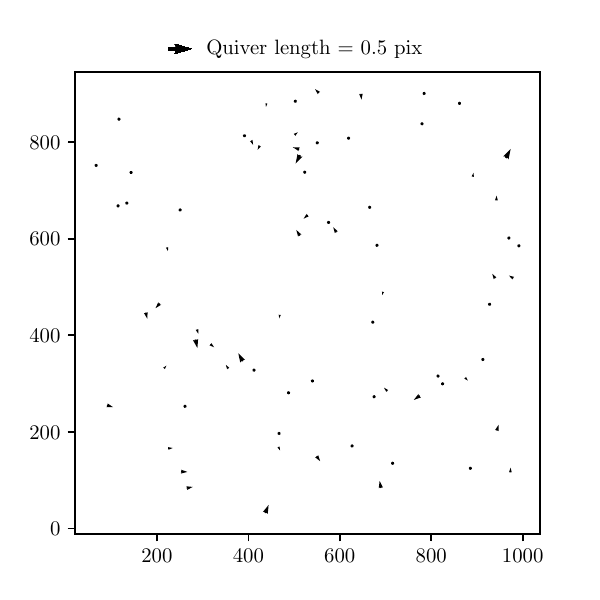}{0.45\linewidth}{}}
\caption{Histogram of $xy$ offsets and quiver plot showing the difference between distortion-corrected 
pixel positions of stars common to both the 2020 and 2021 PyWFS datasets when correcting for an $xy$ shift between the two. We can see that an $xy$ shift alone explains all the discrepancy between these two datasets to $<0.1$ pixel accuracy, indicating that astrometric measurements with NIRC2+PyWFS are repeatable.}
\label{fig:2020to2021}
\end{figure*}
\begin{figure*}
\gridline{\fig{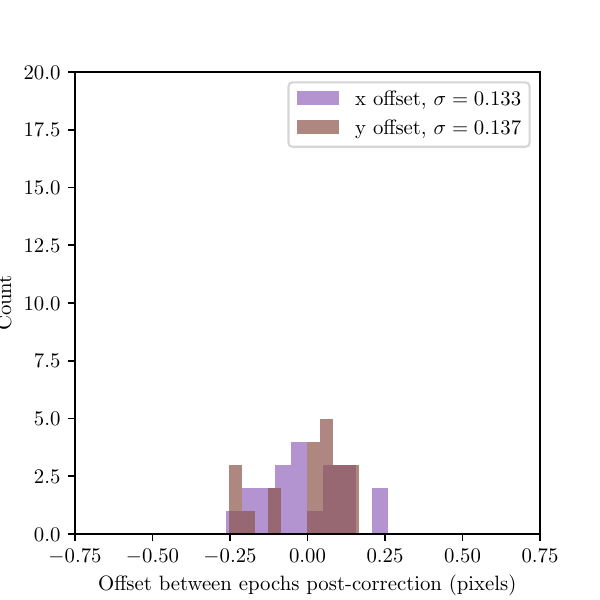}{0.45\linewidth}{}
          \fig{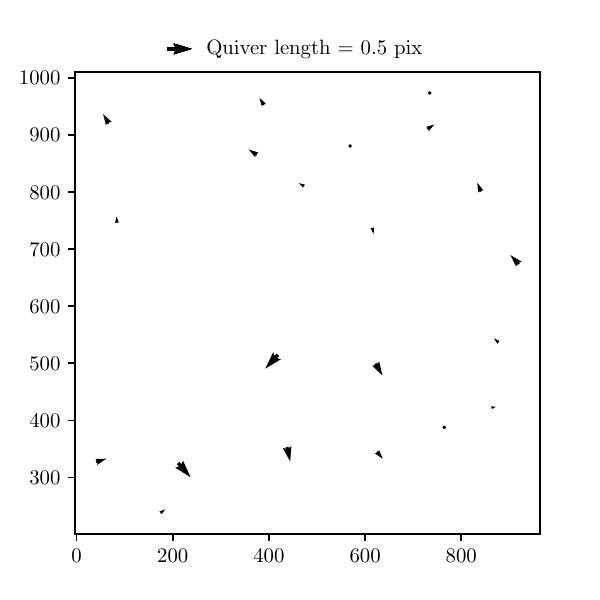}{0.45\linewidth}{}}
\caption{Histogram of $xy$ offsets and quiver plot showing the difference between distortion-corrected 
pixel positions of stars common to both the 2020 and 2022 PyWFS datasets when correcting for an $xy$ shift between the two. The 2022 data were at lower signal-to-noise than other epochs, hence the fewer detected stars. We can see that an $xy$ shift alone explains the discrepancy between these two datasets to $<0.15$ pixel accuracy, showing that these two epochs are comparable.}
\label{fig:2020to2022}
\end{figure*}

When we fit for the transformation from the \citet{yelda2010} data to our May 2020 PyWFS data, we found a rotation angle of $0.118^\circ \pm 0.006^\circ$ anticlockwise (i.e. East of North) and a scale factor of $0.9970 \pm 0.0001$ (shown in Figure \ref{fig:yeldato2020}). This decrease in pixel-coordinate distance between objects corresponds well with the increase in pixel scale from the \citet{yelda2010} correction to the \citet{service2016} correction, as $0.9970 \pm 0.0001$ is consistent within 3$\sigma$ with $0.9981 \pm 0.0004$, the ratio between the \citeauthor{yelda2010} and \citeauthor{service2016} pixelscales. However, the rotational offset of $0.118^\circ \pm 0.006^\circ$ that we find has not been seen before, as \citet{service2016} reports a detector orientation consistent with the \citet{yelda2010} solution. In order to understand if this rotation occurs over time, we compared our 2019, 2021 and 2022 PyWFS data to our 2020 PyWFS data following the approach described above. We found that the differences between our 2020 dataset and the 2021 and 2022 datasets were sufficiently described by a simple translational shift (Figures \ref{fig:2020to2021} and \ref{fig:2020to2022}), and the differences between our 2019 and 2020 data were best described by a combination of a translational shift and a very slight rotation which is within the error budget of the \citeauthor{service2016} solution ($0.025^\circ \pm 0.002^\circ$; see Figure \ref{fig:2019vs2020}). Based on this, we conclude that the change in PA from the time of the \citet{yelda2010} data collection to that of the PyWFS data collection is a global offset, which we correct for in our astrometry pipeline by introducing an additional rotation of $0.118^\circ \pm 0.006^\circ$ clockwise. The additional observed rotation between 2019 and 2020 PyWFS epochs would introduce $\lesssim0.3$ pixels of error into recovered companion positions when using the NIRC2 narrow camera, so we add this rotation in quadrature to our astrometric errors in our analysis.

One factor that we are unable to determine from our data is whether the rotation between the \citet{yelda2010} and PyWFS datasets is induced specifically when observing with the PyWFS, or if it is the result of a change that occurred between the collection of the \citet{service2016} dataset in 2015 (which reports the same PA offset as that of the \citeauthor{yelda2010} data) and the collection of our 2019 PyWFS data. If the former, this effect would have to be accounted for when carrying out astrometric analyses with a combination of contemporary Shack-Hartmann and PyWFS data, and thus might affect our astrometry for XEST 13-010. We see no evidence of an offset between SHWFS and PyWFS in our data, but cannot rule this out.
We also note that we only perform this analysis using NIRC2's narrow camera. Some of our astrometric datapoints derive from images taken with NIRC2's wide camera, though we expect that the additional rotation we see to be camera-independent. Thus, we apply our findings to our wide camera data too. An additional point of note is that it is unlikely that a single dichroic could cause a rotational shift, which is another reason to suspect that this offset is global. As a result of this, we adopt this rotation shift across data obtained with both the SHWFS and PyWFS.


The error in the positions of each object comes from two sources: the astrometric error resulting from the finite signal-to-noise of the data, and any intrinsic residual uncertainty. In order to estimate how large the latter is, we measured the positions of the objects in each individual frame in our 2020 and 2021 PyWFS image stacks, and adopted the standard error for each set of $x$ and $y$ values as the astrometric measurement error. We then constructed a version of the $\chi^2_\nu$ statistic involving a combination of the astrometric error $(\sigma_i)$ and an extra intrinsic error $(\sigma_{\rm int})$. Thus,

\begin{equation}
    \chi^2_\nu \equiv \frac{\chi^2}{\nu} = \frac{1}{\nu} \sum_{i}\frac{(O_i - E_i)^2}{\sigma_i^2 + \sigma_{\rm int}^2},
\end{equation}
where $O_i$ is the observed 2021 PyWFS position measurement, $E_i$ the transformed 2020 PyWFS position measurement, and the error term $\sigma_i$ the quadratic sum of the errors on our 2020 and 2021 PyWFS measurements. The number of degrees of freedom is $N_{\rm stars} - 2 = 86$. We then calculated the value of $\sigma_{\rm int}$ that minimised $\chi^2_\nu - 1$.
We found that $\sigma_{\rm int} = 0.0006$ pixels produced the closest $\chi^2_\nu$ to 1.
We therefore consider this intrinsic error term too small to be relevant to our analysis, and do not account for it in the main paper.

\begin{deluxetable*}{lccccccc}
\centering

\tabletypesize{\small}
\tablecaption{Comparison of 2- vs. 4-parameter fits}\label{tab:distortioncalcs}
\tablehead{\colhead{Epochs} & \colhead{\# stars} &\colhead{MAD in $x$} & \colhead{MAD in $y$} &  \colhead{MAD in $x$}& \colhead{MAD in $y$}  & \colhead{Rotation angle} & \colhead{Scale factor}\\
\colhead{} & \colhead{} & \colhead{(2-param)}& \colhead{(2-param)} &  \colhead{(4-param)}&  \colhead{(4-param)} & \colhead{(degrees)} & \colhead{}}
\startdata
Yelda vs. 2020 & 106 & 0.652 & 0.901 & 0.211 & 0.228 & \phantom{$-$}$0.118^\circ \pm 0.006^\circ$ & $0.9970 \pm 0.0001$ \\
2019 vs. 2020 & 89 & 0.130 & 0.086 & 0.053 & 0.057 & \phantom{$-$}$0.025^\circ \pm 0.002^\circ$ & $0.99997 \pm 0.00004$\\
2020 vs. 2021 & 88 & 0.048 & 0.062 & 0.049 & 0.061 & $-0.005^\circ \pm 0.001^\circ$ & $1.00005 \pm 0.00002 $ \\
2020 vs. 2022 & 29 & 0.109 & 0.111 & 0.173 & 0.130 & $-0.006^\circ \pm 0.006^\circ$ & $0.9997 \pm 0.0002$ \\
\enddata
\tablecomments{The MAD in $x$ and $y$ quoted is the mean absolute deviation of the fit residuals in the $x$ and $y$ directions respectively. We quote the best-fit rotation angle and scale factor from the 4-parameter model, with errors estimated using bootstrap resampling. Rotation angle is defined as the anticlockwise (East of North) angle required to rotate the stars from epoch 1 to match epoch 2. Scale factor is the factor we multiply the coordinates of epoch 1 by to match epoch 2.}
\end{deluxetable*}


\FloatBarrier

\bibliography{bib}{}
\bibliographystyle{aasjournal}



\end{document}